# Emergence of consensus as a modular-to-nested transition in communication dynamics

Javier Borge-Holthoefer[1,2,3]*, Raquel A. Baños[3], Carlos Gracia-Lázaro[3], Yamir Moreno[3,4,5]*

[1] Qatar Computing Research Institute, HBKU, Doha, Qatar
[2] Internet Interdisciplinary Institute (IN3), Universitat Oberta de Catalunya (UOC), Barcelona, Spain
[3] Institute for Biocomputation and Physics of Complex Systems (BIFI), Universidad de Zaragoza, 50018 Zaragoza, Spain
[4] Department of Theoretical Physics, Faculty of Sciences, Universidad de Zaragoza, Zaragoza 50009, Spain
[5] Institute for Scientific Interchange (ISI), Torino, Italy

* To whom correspondence should be addressed: borge.holthoefer@gmail.com, yamir.moreno@gmail.com

Online social networks have transformed the way in which humans communicate and interact, leading to a new information ecosystem where people send and receive information through multiple channels, including traditional communication media. Despite many attempts to characterize the structure and dynamics of these techno-social systems, little is known about fundamental aspects such as how collective attention arises and what determines the information life-cycle. Current approaches to these problems either focus on human temporal dynamics or on semiotic dynamics. In addition, as recently shown, information ecosystems are highly competitive, with humans *and* memes striving for scarce resources –visibility and attention, respectively. Inspired by similar problems in ecology, here we develop a methodology that allows to cast all the previous aspects into a compact framework and to characterize, using microblogging data, information-driven systems as mutualistic networks. Our results show that collective attention around a topic is reached when the user-meme network self-adapts from a modular to a nested structure, which ultimately allows minimizing competition and attaining consensus. Beyond a sociological interpretation, we explore such resemblance to natural mutualistic communities *via* well-known dynamics of ecological systems.

## Introduction

Nowadays, online social networks constitute mainstream ways to communicate, exchange opinions, and reach consensus [1-4]. They are characterized by a multichannel information flow and by an adaptive topology. In recent years, it has increasingly become evident that competition significantly shapes the topology of and the dynamics on these information-driven platforms [5,6], also at the *macro* scale [7]. Given the many sources of information to which a typical individual is exposed, it is likely that the economy of attention rules the system dynamics [8]: even opinion-aligned individuals compete to increase their visibility among other peers, given the limitations of our social brain [5,6]. Such competition may not be direct, but rather mediated by the symbols (memes) that take part in the communicative interaction [9,10] –which similarly compete [11] for the attention of those who produce and consume them (users).

The accent on *intra-class* (user-user, meme-meme) competition renders however a partial picture. Turning to *inter-class* interactions, these appear under the form of mutualism: the choice of more frequent memes increases the visibility of individuals, which makes the popularity of those memes even larger, thus decreasing the likelihood that other competing memes also become fashionable. Under this diversity of actors and connections, an information-driven system can be thought of as a bipartite network in which individuals and memes concurrently compete (within their class) and cooperate (between classes), see Figure 1. Such a system is reminiscent of those that have been reported in other areas, be them plant-animal [12-14] or manufacturer-contractor networks [15], in which nestedness –a widely reported structural pattern in mutualistic ecological systems– is a prominent topological feature. The question is then whether similarities at the dynamical level (same type of interactions) are mirrored at the structural one, and (if so) why a nested architecture, in which specialists –interacting with only a few partners– tend to be connected with generalists –those interacting with many others, emerges.

[Figure 1 around here]

Microblogging platforms stand out as the perfect test bed to answer such question, since messages are explicitly limited to a small number of characters –competition in such restricted environments is fierce, and the choice of memes (hashtags in Twitter, for instance) critically determines the success of the message (outreach) and its lifetime on the system (persistence). Moreover, even though the finding of a nested architecture in bipartite communication networks would be suggestive, online social networks additionally provide us with time-resolved data, which makes it possible to trace back the origins of the nested pattern –at variance with all previous works: we have scarce evidence of how nestedness arises in nature, given the observational limitations and costs of fieldwork [16]. This is the reason why ecologists have focused rather on other aspects [17-19], letting aside the temporal dimension, i.e., the growth and evolution of the system and the emergence of nested patterns.

Here we show that in the information ecology context, it is possible to monitor the emergence of a nested architecture out of an incipient system, which, surprisingly enough first appears under the dominant form of a modular network. To do so, we represent a communication platform as a bipartite graph where connections exist only between agents (users) and the symbols (memes) they produce. Exploiting the inherent time-stamped nature of the data, the bipartite setting yields longitudinal observation of initially modular-and-nested, then nested-only structures from large, public collections of online microblogging data. Additionally, we perform extensive numerical simulations on synthetic networks and find that the observed modular-to-nested transition is due to the fact that the user-meme community is pushed towards a nested architecture to accommodate mutualistic interactions –as opposed to antagonistic ones intrinsic to a modular scenario [20-22]. Our results provide a novel mechanism to explain the emergence of consensus in social systems, and clear the path for a new set of concepts and tools –borrowed from ecology– to be applied in such systems. Last, but not least, our observation of an *empirical* modular-to-nested structural transition can shed light into the problem on the origin of nested architectures, which remains an elusive question.

## Results

We first present results for a dataset corresponding to civil protests in Spain (15M movement) that resonated on Twitter, in the period April-May 2011 [23,24]. The dataset was obtained from a predefined set of keywords relevant to the movement (section A and Appendix of the Supplementary Materials (*SM*) describe in depth all datasets used in the work). These data, taken in *w*-wide sliding windows, contain all the necessary information to build time-resolved bipartite networks –*who* said *what*, and *when*– suitably encoded as a rectangular, time-dependent matrix. Specifically, the Twitter stream is parsed and bipartite graphs –see Figure 1a and 1b– are built up as follows: first, time windows are set to a fixed, arbitrary, $w = t_2 - t_1$ width. We then choose the $n$ most active users and the $m$ memes (hashtags) that those users produced within that time interval. This bipartite network is encoded in an $n \times m$ rectangular binary matrix, $\mathbf{M}_t$, where $t$ indicates the origin of the time window $w$ and $M_{u,h} = 1$ if user $u$ mentioned the hashtag $h$ within the period spanning from $t_1$ to $t_2$ and zero otherwise. This procedure allows generating bipartite networks as time goes on by using a rolling-window scheme to evaluate the evolution of the system, such that a window at time $t$ has a $\varphi w$ overlap with that at time $t - w$ ($\varphi = 0.5$ in the results reported here; for $\varphi$ closer to 1.0 results com at higher resolution, whereas $\varphi = 0.0$ implies non-overlapping windows).

Once the networks associated to the 15M social movement at different times are assembled, we proceed to analyze their structure focusing on two topological characteristics. As the interest is in inspecting whether groups of individuals using the same memes build up, we first look for the optimal modular partition of the nodes through a community detection analysis [25,26], applying a simulated annealing heuristics to maximize Barber's [26] modularity $Q$.

Next, we study whether nested patterns arise in the system. Here we evaluate nestedness following the findings by Bell *et al*. [28,29] and further developed in Staniczenko *et al*. [30], who showed that it is given by the maximum eigenvalue of the $(n+m) \times (n+m)$ adjacency matrix of the network, i.e. the square matrix counterpart of $\mathbf{M}_t$. As shown in Figure S2, our results are robust against other existing measures of nestedness (i.e., NODF [31]). For details on both $Q$ and nestedness, see *Materials and Methods*, and Sections B and C in the *SM*.

Figure 2 shows the results of the application of these structural analyses for the 15M dataset and a window width of $w = 1$ day. If we focus on the days around which the main demonstrations happened (May 15th and onwards), we see that the network presents a highly nested profile. This alone is a quite interesting result, as it implies that when the activity around certain topic peaks, the user-meme system is highly nested. Note that this scenario is more optimal for information diffusion than a predominantly modular topology, as in the latter architecture information flow can get stuck and never reach throughout the whole system. Thus, our findings contribute yet another example of commonalities between ecological, human [15,32] and proto-cultural [33] systems –for which we typically have static perspectives (but see [34,35,46]).

[Figure 2 around here]

Importantly, we can trace back in time the emergence of the final nested state by inspecting the structure of the matrix $\mathbf{M}_t$ at different times $t$. With few exceptions, from the very the beginning of the observation time (April 25, 2011) the network exhibits significant ($z_Q > 1.96$) modularity and nestedness ($z_\lambda > 1.96$) values. This means that before the general onset of collective attention around the 15M activity, the (proto-) topic is composed of a set of modules (Fig. 2, bottom left) which hardly interact with the rest of the system. At the same time, the structure of the network is nested (Fig. 2, top left). Both patterns exhibit a coupled growing trend ($r = 0.7997$) for some time, suggesting that discussion communities become clearer and more internally organized. This picture however changes as the movement gains momentum and consensus arises. Indeed, around the climax of the event (May 15-17) we observe an abrupt transition, i.e., nestedness keeps increasing as modularity collapses in a marked anti-correlated pattern ($r = -0.7819$). After such transition, the architecture of the network is radically different.

The compelling evidence of nested patterns provides a parsimonious explanation of how large amounts of activity can coexist with natural constraints to attention and memory. The user-meme network self-organizes towards a nested structure minimizing competition and facilitating the coexistence of individual participants [36]. Even when the network is predominantly modular, nestedness appears to have significant values well beyond random counterparts, which already indicates the existence of an incipient consensus around sub-topics. Moreover, the unraveled structural change to a highly nested-only architecture allows interpreting the evolution of the Spanish mobilization episodes as a build-up effort from segregation (scattered activists acting locally) to coordination (a global movement with a well-defined and shared main message).

Such interpretation in sociological terms can be quantitatively supported if we actually explore the survival conditions under which the topic can persist. To do so, we build a set of synthetic networks that purposefully present an almost perfect modular architecture, and an almost perfect nested structure (see section E.1 in the *SM* for details), mimicking the initial and "climax" state of the real system, April 25-30 and May 15-20 respectively. To each pair (equal size and equal link density) of these networks, we apply the mutualistic dynamics proposed by Bastolla *et al*. [36] exploring a wide range of model parameters' values (see *Materials and Methods* and Section E.2 in *SM*). The aim is to compare the persistence of these two distinct topologies when equilibrium is reached. The first noticeable finding shows that the nested architecture presents large areas in the parameter space for which the system largely survives, whereas the modular structure does not (Fig. 3a). In all the cases (see additional results in Figs. S7 to S11) it is possible (and actually very frequent) to observe high persistence for the nested architecture whereas it is low for the modular one, but never the other way around. In this context, the persistence is defined as the survival of a hashtag or user once the system has become stable, while the survival rate represents the final diversity (*i.e.*, number of users and hashtags in the steady state) relative to the initial collection. Then, the survival area represents the region with a survival rate greater than a given value (see Section E.2 of the *SM*). We systematically compare the survival areas for pairs of systems with different sizes and densities (Fig. 3b) and two remarkable facts stand out: first, nested architectures consistently out-survive modular ones. Second, the difference in survival areas increases with network size, being narrower for small system sizes. This latter finding suggests the reason why topic-centered bipartite networks in information systems exhibit a modular structure while they remain small-sized: the pressure for an architecture shift remains low, as the transition towards a nested topology does not yet present a critical advantage in terms of the survival of the topic. In other words, when a topic is emerging, and thus its user-meme network is small, it needs to reach a critical mass (here the size) and self-adapt to a nested architecture to increase the likelihood of topic's survival.

[Figure 3 around here]

It is possible to get further insights into the microscopic mechanisms behind the modular-to-nested topological transition. As seen from Fig. 2, once the nested patterns begin to dominate the network structure –around the day when the movement fully develops–, nestedness remains at high levels for some time. This makes it possible to consistently track the set of users and memes that accumulate many interactions (generalists) and inspect whether these sets are time-independent. To this end, we identify which nodes and which memes assemble the core [37,38] of the network at different times. The core can be thought of as the set of most generalist nodes (users and memes) in the network, see section F of the *SM* for further details. Figure 4 compares the resemblance to the "reference core" $D_{RC}$, i.e. similarity between a snapshot's core ($C_t$) and the one extracted when the nestedness is maximal ($C_{max}$) (see section F in the *SM* for a definition). Notably, for both $w$ = 12h (top panel) and $w$ = 3 days (bottom) there is a high turnover in users who occupy the core: in most snapshots $t$, only 0-10% of the users in $C_{max}$ are also present in $C_t$, even when the network's architecture has reached the nested stage. Instead, hashtags have a much more stable core –around 20% of the $C_{max}$ is shared during the entire observation window, and values above 50% are reached after the movement onset and beyond. These results suggest that it is the set of generalist memes, rather than the existence of generalist individuals, that takes the burden of the topic's persistence in time. Indeed, it is less costly to linger on a set of hashtags –the *passive* elements of the system [39]– as they are not subject to users' limitations (sleep, attention focus, etc.), with high volatility of new users who enter and leave the core rather intermittently. As shown in Figure S4, these results are robust to different window widths.

[Figure 4 around here]

Finally, to rule out the possibility that our results are specific to socio-political phenomena of the kind of the 15M movement, we have analyzed an unfiltered dataset of Twitter traffic corresponding to tweets in United Kingdom. As before, bipartite user-hashtag networks are built, but now we chose the subset comprising the top 1,024 most-active users and, independently, the subset of 1,024 most-used hashtags. Note that such independent sampling implies that the corresponding adjacency matrix could be empty – the most active users might not use the most popular hashtags. Results for this dataset show strongly fluctuating patterns for both modularity and nestedness, when measured at large window widths ($w >$ 3h) –not resembling the more persistent, smoothly developed 15M movement. This is not surprising, as most online topics that succeed in getting collective attention do not demand for days to brew and emerge, but they arise and decay at very fast time scales [4, 55]. Figure 5 thus shows the results obtained for the UK dataset over a much shorter time scale ($w$ = 1h), revealing that collective attention around certain topics is reached when the network is maximally nested and minimally modular (with overall $r$ = –0.7126). Here we do not observe coupled modularity-nestedness regimes ($r > 0$), as the incipient stages of a forming topic go unnoticed in the unfiltered stream. For example, a *post hoc* inspection of the unfiltered stream revealed the consolidation (but not the incipient stages) of the XLVIII Super Bowl topic, that started on February 3rd, 2013 at 12:30AM CET, showing the highest peak (lowest valley) in the nestedness (modularity) values in the studied period.

[Figure 5 around here]

## Discussion

In summary, our analyses have unveiled the mechanisms underlying the evolution of an information ecosystem, revealing that there is a traceable pattern for an emerging collective attention event to culminate. Such pattern implies a sudden transition from an initially disperse scenario (modular architecture) to a cohesive situation (nested architecture). Extensive numerical simulations reveal that user-meme mutualistic interactions [9] drive the networked structure towards that nested-only stage, i.e. the architecture that best accommodates the coexistence of individual participants [12,36]. These results stem from an integrated view of the temporal dynamics of emergent collective attention in the context of interdependence and coevolution of human-meme ecosystems, both in online and offline communication.

The present work thus places the study of user-meme structures within the framework of mutualistic communities. This implies that the lessons from such rich tradition can be applied in this new informational context, with the advantage of the finest temporal resolution –time-resolved datasets are scarce in the ecological literature. For instance, the concepts of competition, cooperation and facilitation, vaguely used in reference to information environments, can now be put on firm theoretical standpoints. By connecting meme-mediated human interaction to one of the landmarks in systems ecology –nestedness–, we enlarge the list of complex systems for which such configuration has been reported –with the implications it bears. Such is the case of organizational networks [14,16] or cultural assemblages [33]. Our findings support the idea that nestedness is indeed a dominant pattern in complex networked systems –but it has, paradoxically, received much less attention than modularity. Last but not least, our results provide empirical evidence –at least in the human communication scene– that modularity and nestedness, two dominant architectural principles in complexity, can coexist in a single topology at its early stages, but abruptly bifurcate as the system reaches maturity. Such findings have deep implications on a system, affecting its dynamical properties in terms of diversity, stability, diffusion, and so on. This is then a valuable addition to an ongoing debate about modular and/or nested topologies coexistence, which has mainly occurred in the eco- and biological arena [40-42] but also, implicitly, on the theoretical one [18]. Our results suggest deep constraints not yet fully understood about network formation and evolution, which perhaps analytical efforts can disentangle in the future. This opens the path to further studies along the lines explored here. Finally, the phenomenology of the transition described in this work suggests that the methodological approach presented here could be applied to other datasets, provided that there is a brewing period in which consensus is built up as time progresses. As such, it cannot describe situations in which unexpected [4] or exogenous [55] events (lacking precursory activity) suddenly emerge.

## Materials and Methods

**Data.** The analyzed data comes from two disjoint sets of Twitter collections. The data for the Spanish 15M movement were harvested by a startup company (*Cierzo Ltd*.) for a period of 30 days, starting on April 25, 2011. In that period, protests emerged in Spain in the aftermath of the so-called Arab Spring, with a large demonstration on May 15th and strong echoes up to May 22nd (local elections in Spain). Thus our analysis covers a brewing period with low activity rates (up to May 15th) plus an "explosive" phase beyond that day, decaying beyond May 22nd. Our collection comprises 521,707 messages. The UK collection is not filtered topic-wise in any sense. It comprehends almost 29 million messages for a three month period in 2013, the only restriction being the localization of these tweets: they correspond to messages emitted either from the United Kingdom or Ireland. See section A of the *SM* for details on the events and data collection in both cases.

**Bipartite modular structure.** Community analysis is performed *via* Barber's modularity $Q$ maximization. In his work [26], Barber provides an appropriate null model given the bipartite nature of our networks. In particular, a bipartite network is represented as a block off-diagonal binary matrix:

$$A = \begin{bmatrix} O_{n\times n} & \tilde{A}_{n\times m} \\ (\tilde{A})^T{}_{m\times n} & O_{m\times m} \end{bmatrix} \quad (1)$$

and the adequate null model for it is:

$$P = \begin{bmatrix} O_{n\times n} & \tilde{P}_{n\times m} \\ (\tilde{P})^T{}_{m\times n} & O_{m\times m} \end{bmatrix} \quad (2)$$

where $O_{i\times j}$ is the all-zero matrix with $i$ rows and $j$ columns.

All of this is reflected in the magnitude to optimize:

$$Q = \frac{1}{L}\sum_{i=1}^{n}\sum_{j=n+1}^{n+m}\left(\tilde{a}_{ij} - \tilde{p}_{ij}\right)\delta\left(g_i, h_j\right) \quad (3)$$

where $L$ is the number of interactions (links) in the network, $\tilde{a}_{ij}$ denotes the existence of a link between nodes $i$ and $j$, $\tilde{p}_{ij} = k_i k_j / L$ is the probability that a link exists by chance, and $\delta$ is the Kronecker delta function, which takes the value 1 if nodes $i$ and $j$ are in the same community, and 0 otherwise. We give some additional details in Section C of the *SM*. Note that the off-diagonal blocks $\tilde{A}_{n\times m}$ and $(\tilde{A}_{n\times m})^T$ in (1) correspond to $\mathbf{M}_t$ and $(\mathbf{M}_t)^T$ respectively. Modularity z-scores (Figures 2 and 5) have been obtained against the average and standard deviation of an ensemble of $Q$ for $10^2$ random realizations of **A**.

**Nestedness.** In interaction networks, nestedness indicates the extent to which specialists interact with proper nested subsets of those species interacting with generalists [12]. Among many methods to quantify nestedness in bipartite networks, here we evaluate it following the spectral approach [28-30], i.e. the level of nestedness is given by the maximum eigenvalue $\lambda_{max}$ of the adjacency matrix **A** of the network. Nestedness z-scores ($z_\lambda$ *in* Figures 2 and 5) have been obtained against the average and standard deviation of an ensemble of $\lambda_{max}$ for $10^4$ random realizations of **A**. Section B of the *SM* discusses the robustness of the reported results compared to NODF [31], as well as across the most used of significance tests.

**Mutualistic dynamical model.** We model a topic's evolution integrating the set of differential equations in Bastolla *et al*. [36] for both classes of nodes on top of synthetic networks, which have been purposefully built to be almost perfectly modular, and almost perfectly nested (see section E.1 of the *SM* for further details). In particular, we consider a mutualistic community consisting of $n$ users and $m$ different hashtags (memes); the diversity is denoted by $N(t)$ and refers to the sum of active users and hashtags at a given moment $t$. Let $U$ be the set of users and $H$ be the set of hashtags, $s_u$ refers to the relative activity of user $u$ and $s_h$ represents the relative frequency of hashtag $h$. In order to model the evolution of the system, we consider that elements of the same group (users or hashtags) are in competition between each other, while they hold a mutualistic relationship with elements of the other group. Therefore, the activity of a given user $u$ evolves according to [36]:

$$\frac{1}{s_u}\frac{ds_u}{dt} = \alpha_u - \beta \sum_{v \in U} [s_v(\rho + (1-\rho)\delta_{uv})] + \frac{\gamma \sum_{h \in H} m_{uh} s_h}{1 + \lambda\gamma \sum_{h \in H} m_{uh} s_h} , \qquad (4)$$

where the first term $\alpha_u$ represents the specific growing rate. The second term of eq. (4) refers to the competition, where $\delta_{uv}$ is the Kronecker's delta (taking the value 1 when $u = v$ and the value 0 otherwise) and the parameter $\rho$ modulates the strength of the competition between different users (in correspondence with the biological interspecific competition term). Finally, the third term of eq. (4):

$$\frac{\gamma \sum_{h \in H} m_{uh} s_h}{1 + \lambda\gamma \sum_{h \in H} m_{uh} s_h} \qquad (5)$$

models the mutualism. The user-hashtag interactions are represented through the bipartite graph $\mathbf{M}_t = \{m_{uh}\}$, where $m_{uh} = 1$ if user $u$ has posted a message containing the hashtag $h$, and 0 otherwise. $\lambda$ corresponds to the Holling term that imposes a limit to the mutualistic effect, decreasing the mutualistic term to $1 / \lambda$ for large frequencies. The formula for the evolution of hashtags can be obtained by interchanging the indices of the equation:

$$\frac{1}{s_h}\frac{ds_h}{dt} = \alpha_h - \beta \sum_{i \in H} [s_i(\rho + (1-\rho)\delta_{hi})] + \frac{\gamma \sum_{u \in U} m_{uh} s_u}{1 + \lambda\gamma \sum_{u \in U} m_{uh} s_u} \qquad (6)$$

Our aim is to study the evolution of the system in different topologies (nested versus modular) focusing on the survival of memes and hashtags, that is, the diversity in the stationary state. To this end, we performed extensive numerical simulations by integrating the $N$ coupled differential equations (4,6) by means of a fourth-order Runge-Kutta method. Section E.2 in the *SM* reports the explored parameter space and other details.

**Author contributions.** JB-H and YM conceived the original idea for the experiment, JB-H, RAB and CG-L implemented the software for the analysis of data, RAB and CG-L performed the simulations, JB-H, RAB, CG-L and YM analyzed the data and discussed the results, JB-H and YM were the leading writers of the manuscript. JB-H, RAB, CG-L and YM read and approved the final version of the manuscript.

**Acknowledgments.** We thank A. Arenas for comments, suggestions and intense discussions. RAB was partly supported by the FPI program of the Government of Aragón, Spain. Y. M. acknowledges support from the Government of Aragón, Spain through a grant to the group FENOL, by MINECO and FEDER funds (grant FIS2014-55867-P) and by the European Commission FET-Proactive Project Multiplex (grant 317532).

**Data availability.** All data needed to evaluate the conclusions in the paper are present in the paper, the Supplementary Materials and/or http://www.jbh.cat/data.

**Competing financial interests.** The authors declare no competing financial interests.

**Figures**

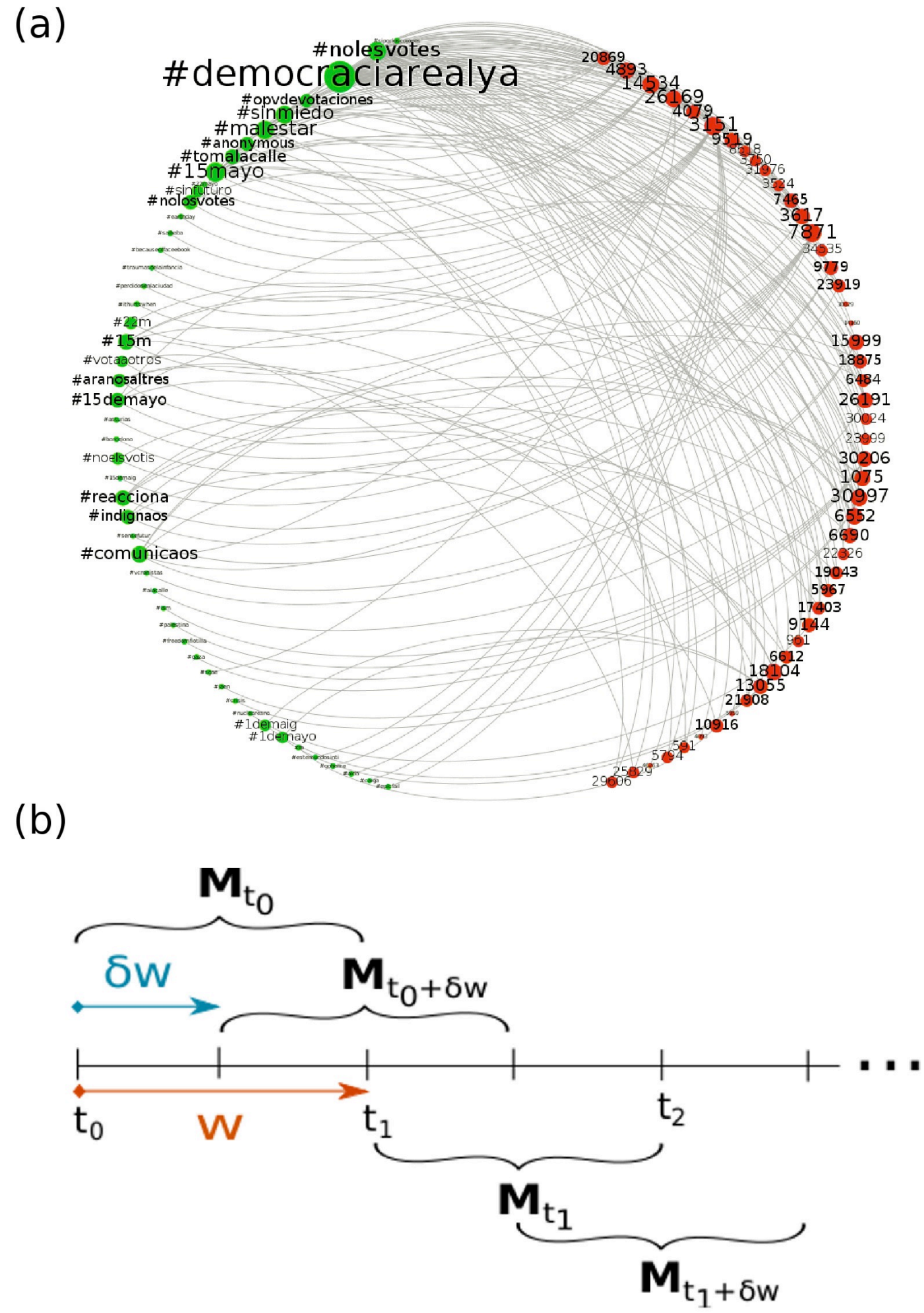

**Fig. 1. Time-dependent Mutualistic Networks.** Bipartite representation of the user-hashtag interaction network at the beginning of the observation period (a). An undirected link is shown whenever a user (represented here by an integer number) authors a tweet containing the corresponding hashtag. The size of hashtags and users is proportional to their frequency/activity. Panel (b) sketches the sliding-window scheme, which produces the matrices $\mathbf{M}_t$ that contain the interactions between users and hashtags starting at time $t_0$ and lasting till time $t$, with $w = t - t_0$.

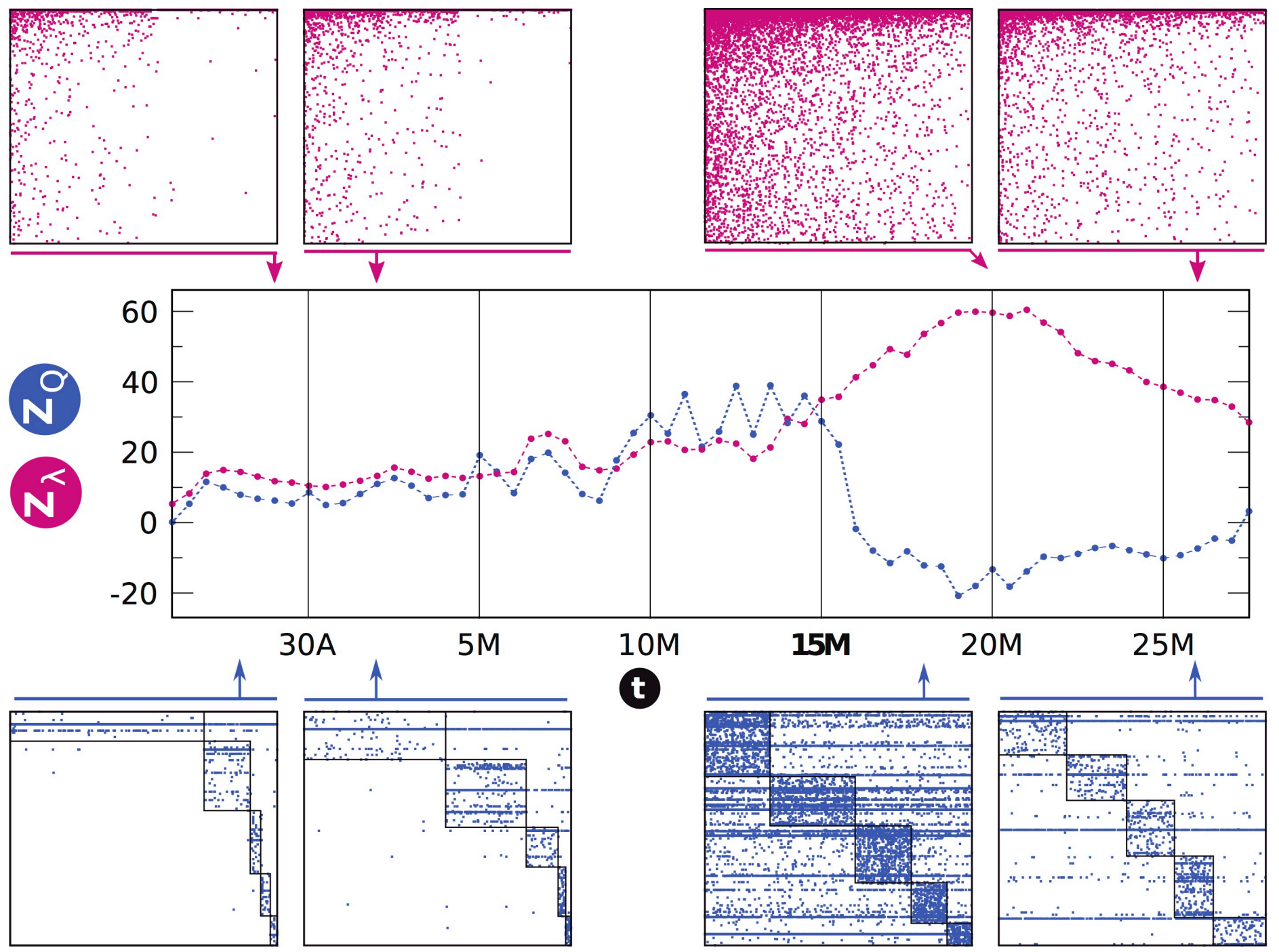

**Fig. 2. Modularity and nestedness bifurcate at the onset of system-wide attention.** The central panel shows the evolution of modularity and nestedness, as standardised z-values. Remarkably, both metrics evolve in a coupled way up to the onset of the main protests (around May 15). At this point, modularity collapses, whereas nestedness continues growing towards its peak value coinciding with the political movement's central dates – that of the largest demonstrations across the country (May 17-20th). Top panels represent four snapshots of the data –encoded as bipartite networks–, rows and columns are sorted in decreasing connectivity order (for an optimal visualization of nested patterns, if they exist). Similarly, lower panels represent the exact same matrices, where rows and columns are sorted module-wise (for an optimal visualization of the community structure).

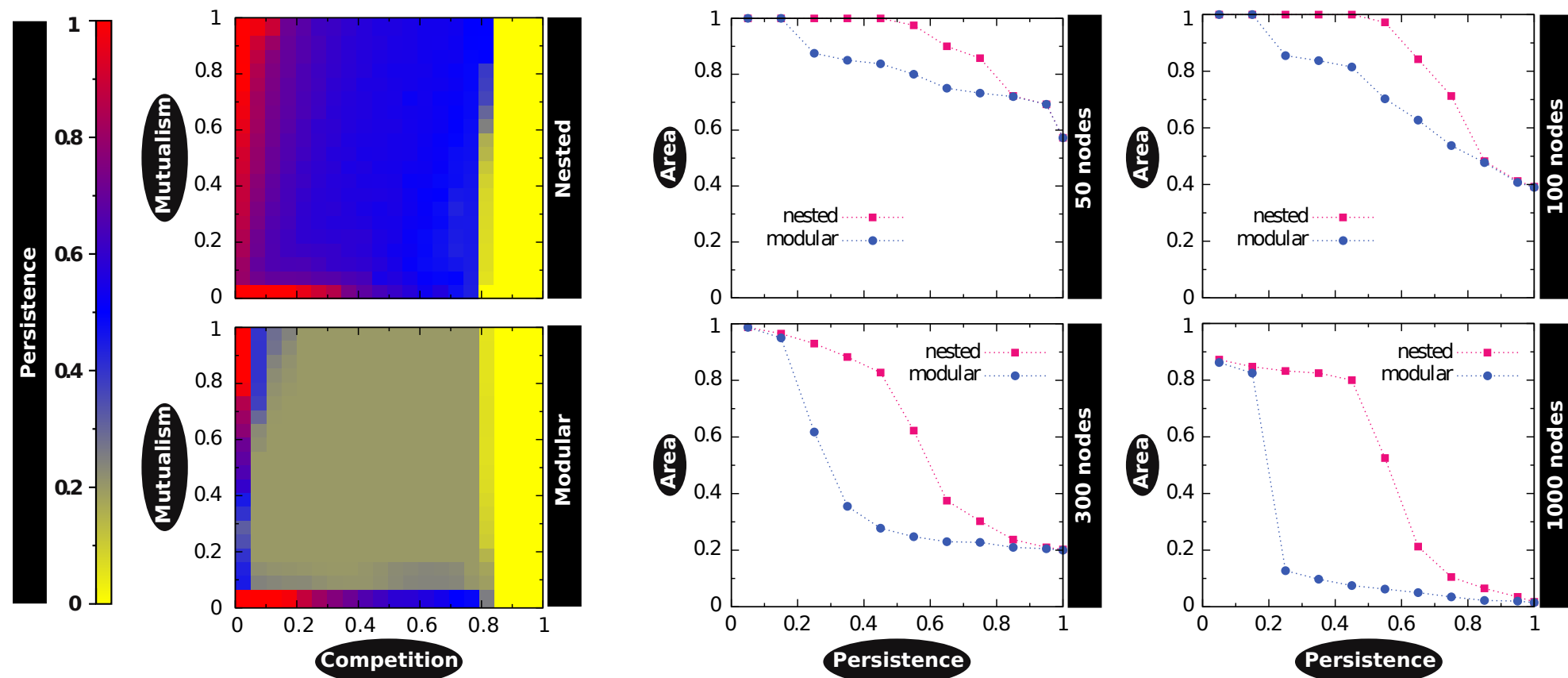

**Fig. 3. Modular and nested architectures under mutualistic dynamics.** Two synthetic networks with the same size ($N = 1000$) and density ($\rho = 0.25$), but different architecture (modular, nested), exhibit radically different outcomes when the mutualistic dynamical framework is applied on them via extensive numerical simulations. Left: Persistence as a function of the competition $\beta$ and mutualism $\gamma$ terms. For a wide range of parameters the modular network shows poor survival; conversely, the nested architecture performs equally or better than the modular counterpart in any given region. Right: differences in the "survival areas" increase with size, which indicates that the pressure for an architectural shift (modular to nested) grows as new nodes (users and hashtags) join the system. Note that the $x$-axis in the right panels ("Persistence") corresponds to the $z$-axis (color code) in the left panels. All results are averaged over 1000 realizations. Additional results for other sizes and densities can be found in Figures S7 to S11.

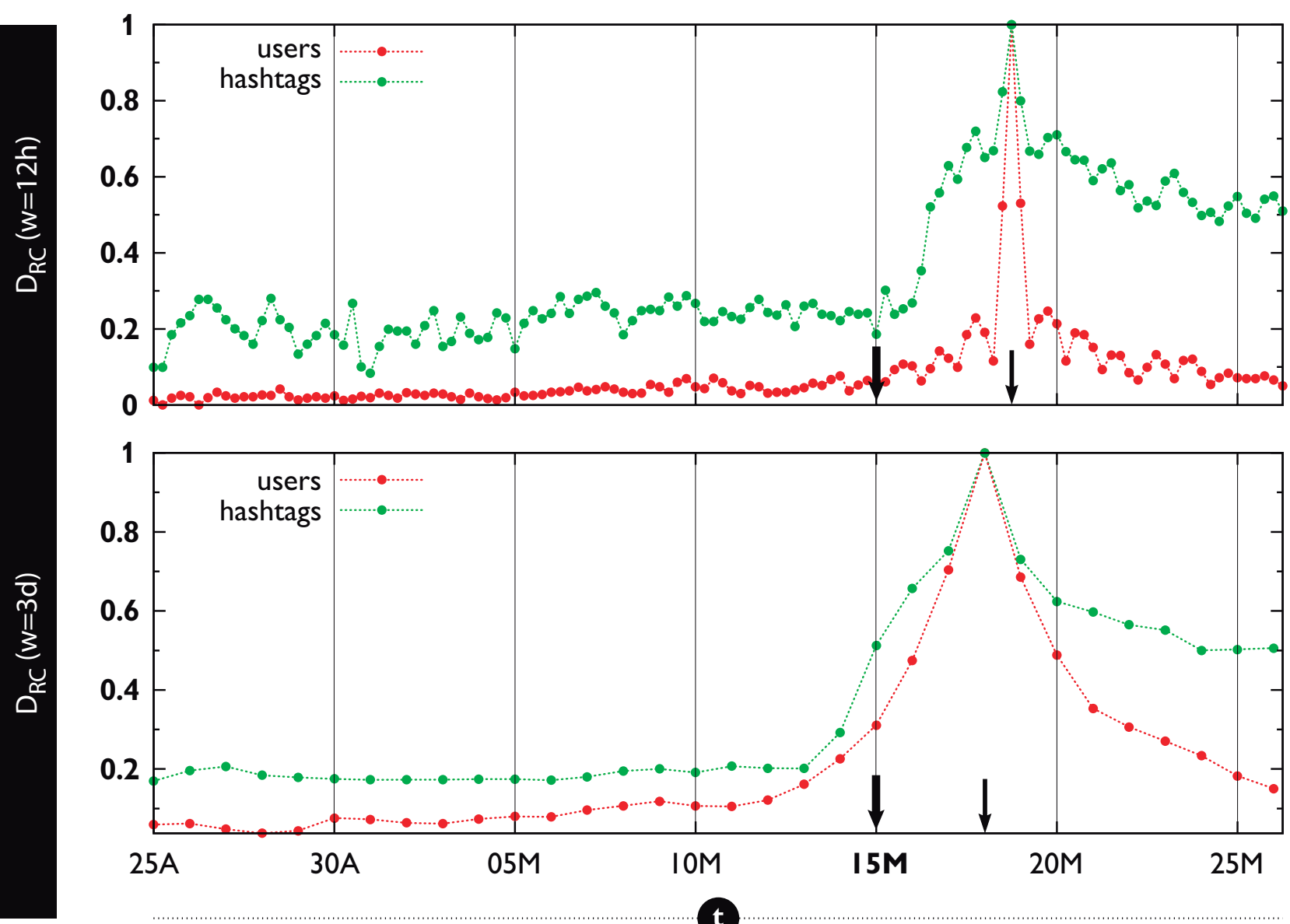

**Fig. 4. Topical consistence over time despite user turnover.** We track the similarity $D_{RC}$ of the generalist cores of the network in time ($C_t$) with respect to a fixed reference (the core of the network when the maximum of the nestedness is observed, $C_{max}$). Results for different $w$ (12h in the top panel; 3 days in the lower) show that only hashtags build a stable core, guaranteeing the semantic coherence of the topic across time; whereas the core of users suffers a high rate of turnover, indicating that users are frequently pushed to and from the periphery of the network.

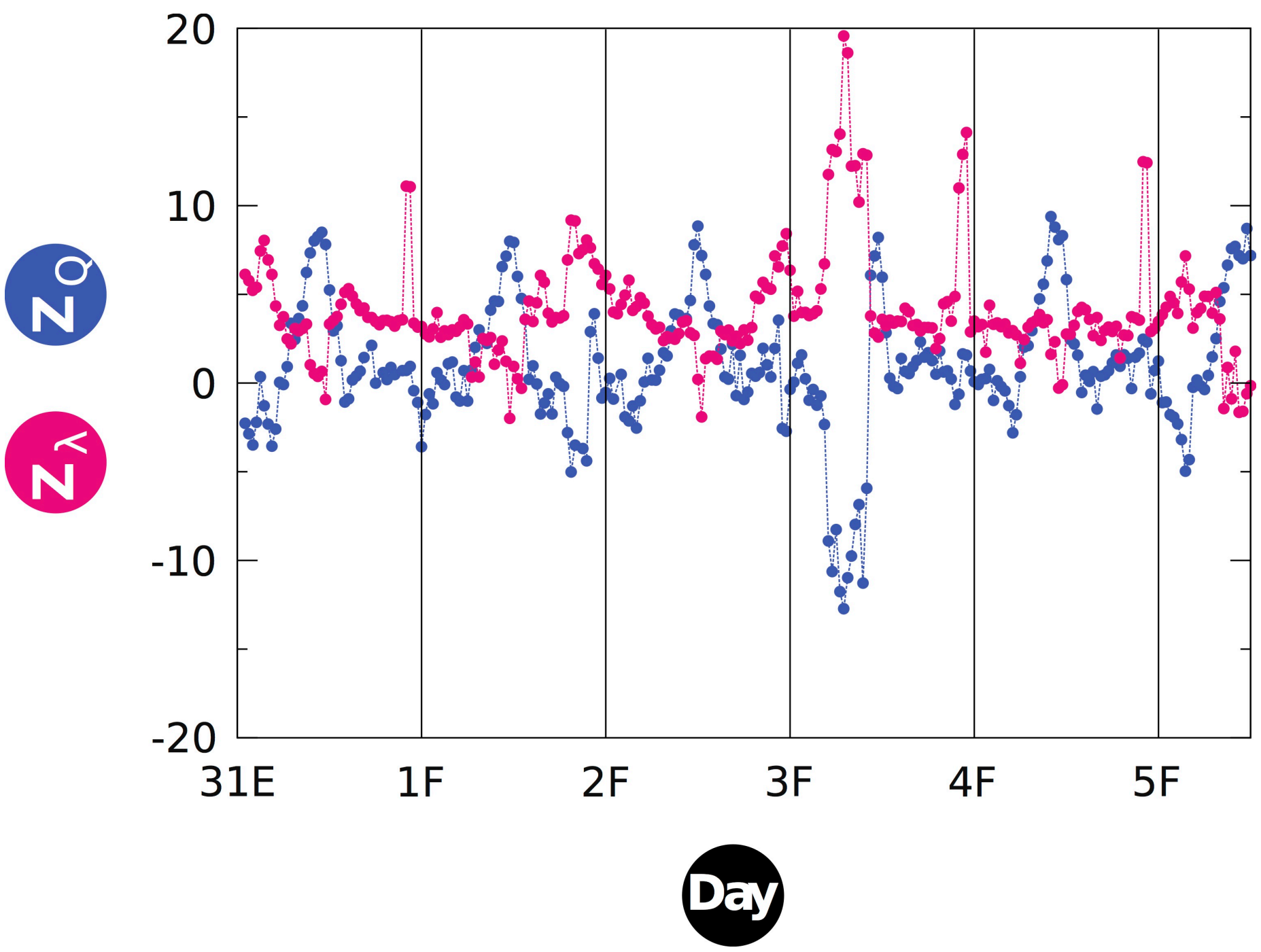

**Fig. 5. Maximum nestedness marks ephemeral topics at faster time-scales.** Unfiltered, topic-independent Twitter traffic offers similar evidence as the main example (Figure 2), provided that a suitable time-scale is examined. In particular, nestedness and modularity show strongly anti-correlated behavior ($r$ = -0.7126), with $z_\lambda$ peaking when a collective attention gathers around an outstanding topic (the most notable one in this plot being the Superbowl event, between February 3 and 4, 2013).

**Supplementary Materials to**

# Emergence of consensus as a modular-to-nested transition in communication dynamics

Javier Borge-Holthoefer[1,2,3]*, Raquel A. Baños[3], Carlos Gracia-Lázaro[3], Yamir Moreno[3,4,5]*

[1] Qatar Computing Research Institute, HBKU, Doha, Qatar
[2] Internet Interdisciplinary Institute (IN3), Universitat Oberta de Catalunya (UOC), Barcelona, Spain
[3] Institute for Biocomputation and Physics of Complex Systems (BIFI), Universidad de Zaragoza, 50018 Zaragoza, Spain
[4] Department of Theoretical Physics, Faculty of Sciences, Universidad de Zaragoza, Zaragoza 50009, Spain
[5] Institute for Scientific Interchange (ISI), Torino, Italy

* To whom correspondence should be addressed: borge.holthoefer@gmail.com, yamir.moreno@gmail.com

## A. Data

The *information ecosystem* under consideration stems from two disjoint sets, which correspond to hashtags and users from the online platform www.twitter.com. Following the analogy with interactions in ecological systems, two types of *species* are considered: users and hashtags (memes). This implies that we do not consider follower/following links between users, nor connections between hashtags due, for example, to co-occurrence in the same tweet. We do not consider either message (tweet) contents, except to extract the hashtags in it: our final aim is to keep record of who-said-what in terms of agents using hashtags.

**Spanish collection**. The collection from Spain comprises 521,707 tweets containing at least one hashtag, 22,375 unique hashtags and 78,080 unique users. The observation period ranges from the 25th of April at 00:03:26 to the 26th of May at 23:59:55, 2011, and data was collected by selecting all the tweets containing at least one of a preselected set of 70 hashtags related to the 15M movement with the aim of filtering out only the activity related with this topic (see Table S3). Data collection was carried out by the start-up Spanish company *Cierzo Development Ltd*.

**UK collection**. The UK collection comprises 28,928,528 tweets emitted by a set of 842,745 unique users between the 18th of January at 18:41:56, to the 31st of May at 22:41:56, 2013. The set of unique hashtags in this case is 2,196,934. Unlike the case of the Spanish dataset, tweets have been filtered by selecting only those that are geolocalized in the United Kingdom and Ireland. In this way, this set provides a raw dataset (only limited by geolocalization) of twitter traffic in which the activity is not filtered by topic, hashtags or users. See Table S4 for a glimpse on the most common hashtags.

### A.1. Data as an evolving bipartite graph

As it is clear in the main text, we attempt to account how the user-hashtag ecosystem changes over time. To this end, we build a sequence of snapshots out of the data. These snapshots have an arbitrary window width $w$, and adjacent snapshots have an overlap of $\varphi w$. Such overlapping scheme is a rather standard procedure when considering chunked temporally-resolved information, to provide a smooth account of change in time.

The question remains how these datasets can be suitably represented. The most natural way to map user-hashtag interactions is through a bipartite graph of relations, which in turn corresponds to a rectangular presence-absence matrix $\mathbf{M}_t = \{m_{uh}\}$, where $m_{uh} = 1$ if user $u$ has posted a message containing $h$, and 0 otherwise (note that matrix $\mathbf{M}_t$ corresponds to block $\tilde{\mathbf{A}}_{n\times m}$ of the block off-diagonal binary matrix $\mathbf{A}$ in eq. (1) of the main text). Noteworthy, this implies that only binary values are considered, i.e. the number of interactions between nodes $u$, $h$ is not recorded. Besides, we acknowledge that results in the main text are not affected by the chosen window width (results there correspond to $w$ = 12 hours and 3 days, with overlaps of 6 and 36 hours, respectively). See below for more details.

It is also important to highlight that the $\mathbf{M}_t$ matrices may not contain the same nodes across $t$: as time advances, users join (disappear) as they start (cease) to show activity; the same applies for hashtags, which might or might not be in the focus of attention of users. This volatile situation is quite normal in time-resolved ecology field studies [34, 46, 52], where the accent is placed on the system's dynamics – rather than individual species. Moreover, the level of turnover in the sequence of data is very informative, as it characterizes how the system renews its structure over time (see the main text).

### A.2 Pruning the data

The large size of our two datasets –78,081 unique users and 22,376 unique hashtags in the 15M dataset, and 842,745 users plus 4,217,530 hashtags in the UK dataset– handicaps the data processing and makes the calculations time-consuming. We must therefore apply some restrictions to the number of users and hashtags considered in the network.

Therefore we apply a rather straightforward criterion, by which we prune the least active users in the data. This means that only top-contributors (and their associated hashtags) show up in the matrices that we study. In doing so, we guarantee that the whole approach makes sense: only by including the most active users we make sure that generalists and specialists will show up –if any nested patterns are to be found. Also the probability of obtaining a connected matrix is higher. Again, we acknowledge that ours is an arbitrary decision. To provide solid evidence, we have tried several matrix sizes.

**Spanish dataset**. Whereas results reported in the main text are based on the 1,024 most active users, we have also tested smaller sets with qualitatively the same results (see Figure S1). In this Figure, we represent the standardized results for both nestedness (left) and modularity (right). Both magnitudes will be described in details later (section B and C). Three dates are also considered at different moments of the 15M movement: three days before the main camps took place –May 12th–, at the onset of the protests; May 15th itself; and May 19th, when the maximum nestedness is achieved and protests are considered to have reached high levels of visibility. Nestedness curves show a tendency to saturate for large values of the number of users selected. This flattening is achieved at lower values for earlier dates, being far from saturation on May 19th. In view of these results, we can safely conclude that our pruning procedure, i.e., the restriction to the most active users, does not give rise to a misleading claim about the nested structure organized around the movement formation. So far, and to avoid extrapolation, we can safely state from Fig. S1 that, if we build a nestedness time series and a modularity time series (admittedly both quite poor: only 3 points each) we see that 19M shows, nestedness-wise, a maximum, regardless of the size one wishes to pick; and modularity-wise a minimum, regardless of the size on wishes to pick (except for very small sizes, $n < 150$). In summary, for any size reported in Fig. S1,

$$z_N^{12M} < z_N^{15M} < z_N^{19M}$$

and

$$Q^{12M} > Q^{15M} > Q^{19M}$$

which would render perfect anti-correlation (that is, a stronger result than the one reported in the main text).

**UK dataset**. For this dataset, the filter is applied in a slightly different way: the cutoff is applied to both users and hashtags, by choosing the 512 more active users and the 512 most-used hashtags. The reason underlying the additional constraint on hashtags and the smaller number of nodes considered, is the large amount of hashtags used in this dataset: 1,024 users can generate from 2,245 to 13,113 hashtags, depending on the observation time window. Some technical details about the observation period, number of users and hashtags, and time-windows width can be found in Table S1.

As for how we build bipartite networks for the UK dataset, different possibilities arise: on the one hand, we could randomly select a subset of users and hashtags involved in the network, but in this way we might be missing the relevant agents thought to play a major role in the contribution to the nestedness of the whole system. Besides, a random selection could lead to empty matrices (none of the selected users tweeted any of the selected hashtags). We must, nevertheless, remark that this situation is highly unlikely for the 15M event, as a result of the very nature of the dataset: only people and hashtags related to this particular topic were extracted from Twitter. We will be making use of this method as a way to compare the nestedness levels in the 15M with a topic null model, built from data from the UK dataset.

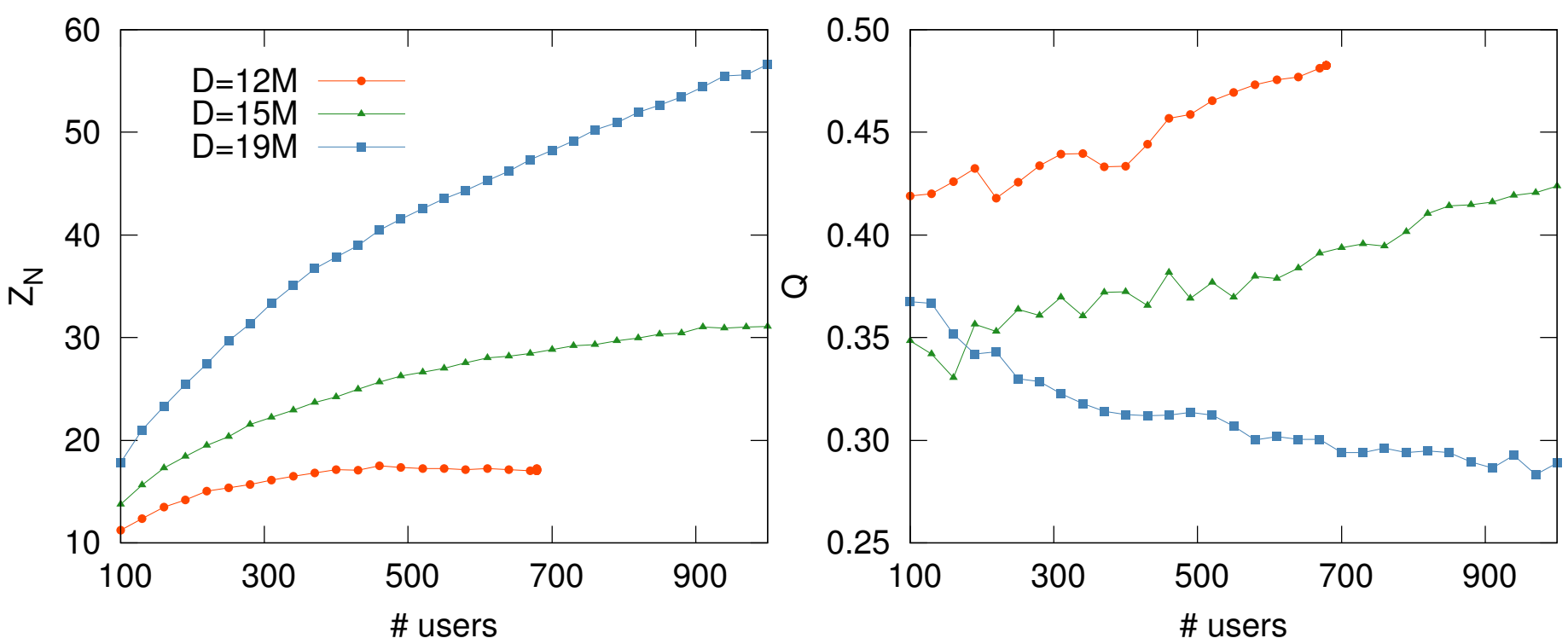

Figure S1: Robustness against matrix size. For the $w$ = 12h set some days have been selected. We perform the null model analysis for different cutoffs in the number of users ($x$ axis), and show how the standardised leading eigenvalue (left) and the standardised modularity (right) evolve. The end at ~700 users for the curve corresponding to $D$ = 12M indicates that the largest possible matrix has been reached, i.e there are no more active users at that particular day. These results not only guarantee that our conclusions about the nested structure around the 15M are robust, but also show that the observed peak would be more prominent if we considered the real matrix including all the users and hashtags.

| | 15M, 2011 | UK, 2013 | UK (inset), 2013 |
|---|---|---|---|
| Date range | 25/04 to 26/05 | 18/01 to 31/05 | 31/01 to 06/02 |
| # total users | 78,081 | 842,745 | 122,553 |
| # total hashtags | 22,376 | 4,217,530 | 264,291 |
| Filter to users | 1,024 | 1,024 | 512 |
| Filter to hashtags | -- | 1,024 | 512 |
| Final # users | 17,202 | 50,091 | 37,174 |
| Final # hashtags | 12,384 | 19,905 | 22,933 |
| Time windows | 12h; 72h | 24h | 60min; 120min |
| Overlap | 6h; 36h | 12h | 30min; 60min |

Table S1: Datasets summary details. The date range, number of total users and hashtags are displayed. We also indicate the cutoff in the number of users and hashtags (if any) that has been applied. An unspecified hashtag filter indicates that the hashtag set is determined by the set of selected users. Users are filtered by activity and hashtags by usage. We also show the final number of users and hashtags after the selection process. Finally, the window width and overlap between consecutive windows are also displayed.

## B. Nestedness in online social networks

**Robustness across metrics**. Several studies have been focused on quantifying nestedness, the first proposals being made by Hultén [48], Darlington [44] and Daubenmire [45] to describe patterns in which species-poor sites are proper subsets of those ones present at species-rich sites. Nestedness analysis has become very popular among ecologists, and, although the concept is widely accepted, it has not been formally defined, yielding to several distinct metrics [30,31,43]. In this work (main text), we adopt a definition numerically confirmed by Staniczenko et al. [30], where nestedness is given by the maximum eigenvalue of the network's adjacency matrix. This metric is based on a theorem regarding chain graphs first provided by Bell et al. [28, 29], where it is shown that among all the connected bipartite graphs with $N$ nodes and $E$ edges, a perfectly nested graph gives the larger spectral radius. The method is advantageous over other possibilities due to the invariance of eigenvalues under matrix permutations, and the remarkably low computation time required to perform eigenvalue calculations, even for large matrices. This is an important detail provided that z-scores for nestedness are obtained against $10^4$ random realizations.

Nevertheless, we have checked the validity of our results against the improved metric NODF, defined by Almeida-Neto *et al*. [31]. This measure is based on two simple properties: decreasing fill (DF) and paired overlap (PO). Assuming that row (column) *i* is located at an upper position in the sorted presence-absence matrix from row (column) *j*, the decreasing fill condition imposes that a pair of rows (columns) can only contribute to the nestedness if the marginal total –the number of interactions a row (column) has– of row (column) *i*, is greater or equal to the marginal total of row (column) *j*. In this case, the paired nestedness, $N_{ij}$, is equal to the paired overlap $PO_{ij}$, i.e., the number of shared interactions between rows (columns) *i*, *j*. The metric can be summarized as:

$$\mathrm{NODF} = \frac{\sum_{ij} N_{ij}}{\frac{m(m-1)}{2} + \frac{n(n-1)}{2}} \qquad (1)$$

where

$$N_{ij} = 0 \quad \text{if} \quad \mathrm{MT}_i < \mathrm{MT}_j$$
$$N_{ij} = \mathrm{PO}_{ij} \quad \text{if} \quad \mathrm{MT}_i \geq \mathrm{MT}_j \qquad (2)$$

Both metrics are compared in Figure S2. In the *x*-axis the standardized value of the leading eigenvector is displayed against the standardized NODF measure. Matrices involved in the plot correspond to graphs at the distinct snapshots with time-window *w* = 1d. These results are displayed along with the Pearson and Spearman coefficients and their *p*-values, showing a good linear correlation with *p*-values $p < 10^{-3}$.

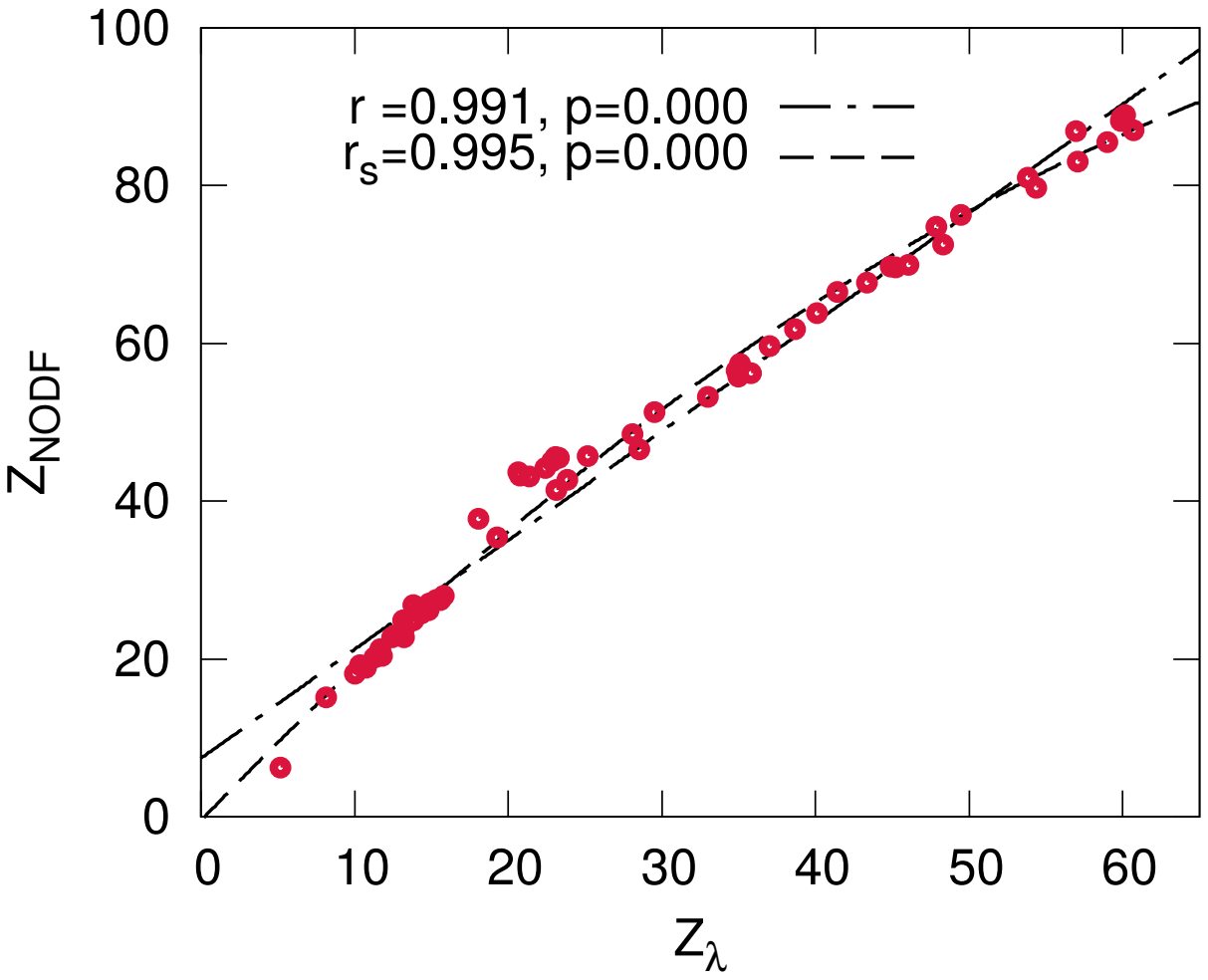

Figure S2: Comparison against nestedness metrics. For every matrix from the set of time windows with *w* = 12h, the standardised leading eigenvalue, $z_\lambda$, and the standardised NODF metric, $z_{NODF}$, are computed. There is a good agreement between both metrics, as the Pearson coefficient, *r*, and the Spearman coefficient, $r_s$, show along with their *p*-values.

**Robustness across significance tests.** The fact that real matrices are usually far from being perfectly nested, imposes the use of a test for the significance of nestedness values. Such a test implies the implementation of a null model and the computation of standardized results, and additionally, allows one to compare matrices with distinct sizes, this comparison being impossible otherwise. Regarding modularity, the metric already includes in its very definition a null model, in such a way that the modularity obtained is already a comparison with a randomized counterpart of the network.

Different null models may be proposed. For example, one could think of a null model rewiring the set of links present in the network. A strict application of such scheme would not maintain the bipartite structure of the network, and for that reason it should be avoided.

Within this restriction we can still think of some variations. Here we explore two different possibilities, as discussed in [40]. In null model I, the number of links in the network is preserved, but placed at random within the matrix –although respecting the class of the origin and end of it. The degree sequence is therefore not preserved. Null model II is a probabilistic null model where an interaction between hashtag $h$ and user $u$ is established with probability proportional to their connectivity,

$$p_{uh} = p_{hu} = \frac{1}{2}\left(\frac{k_h}{m} + \frac{k_u}{n}\right) \tag{3}$$

In the above expression, $n$ stands for the total number of users, i.e., the first dimension of $M_{uh}$, and m for the number of hashtag, equal to the second dimension of $M_{uh}$. $k_u$ and $k_h$ correspond to the degree of user $u$ and hashtag $h$, respectively. This model maintains the number of interactions per class only approximately, i.e. it probabilistically maintains the observed total number of interactions.

We can go further and consider an X-swap scheme null model III, in which a rewiring of the edges is applied but keeping constant the degree sequence of the nodes in the system. This null model, however, is too restrictive, and gives a small number of possible configurations, specially for those matrices having few non-empty cells. We must consider null models having a balance between the number of possible configurations and strictness. For this reason we choose to discard null model III, and apply the probabilistic null model II, which is the strictest between models I and II. Figure S3 reports the consistency of the results for the nestedness using either of the chosen null models. Z-scores have been calculated over 10,000 randomizations.

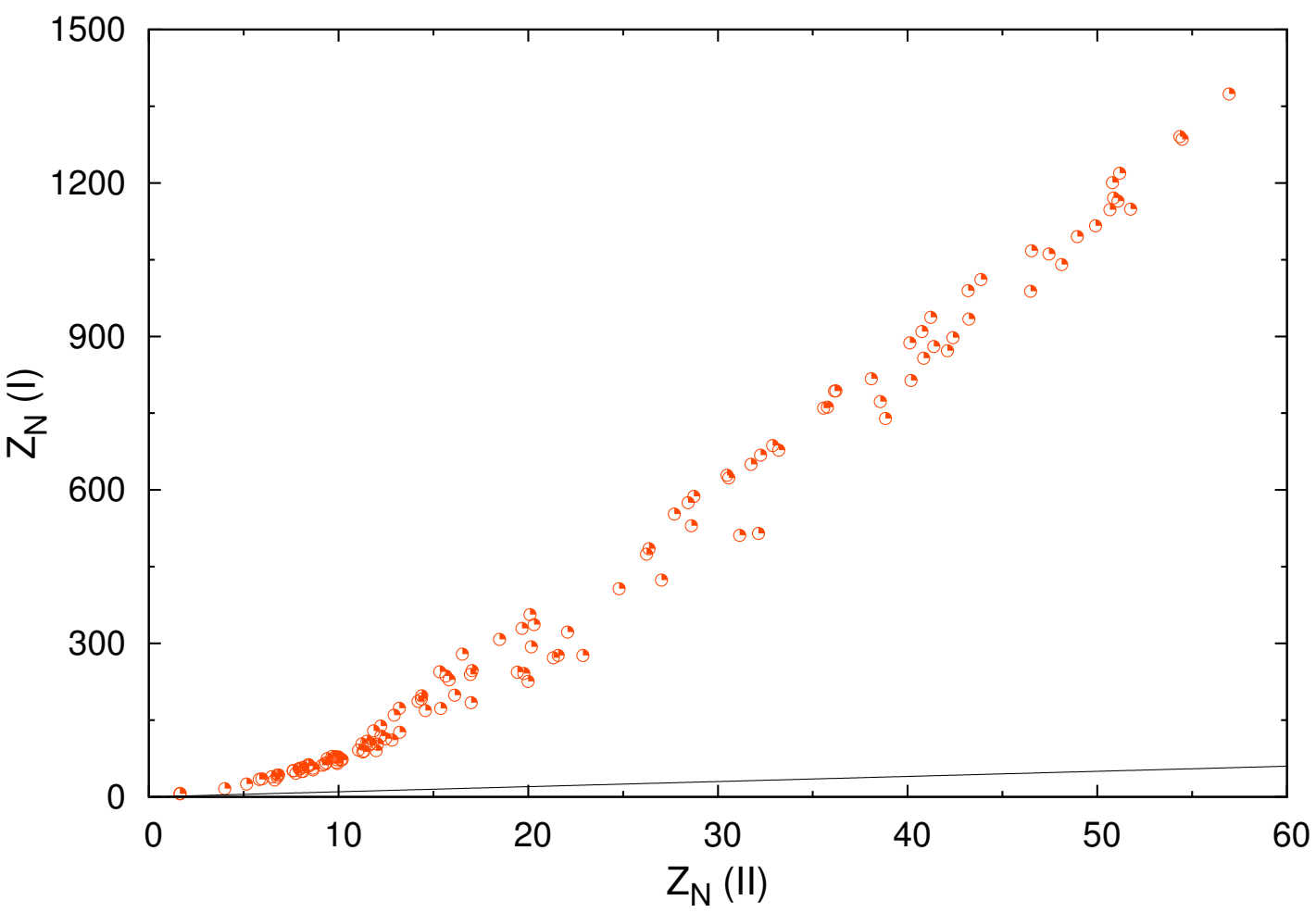

Figure S3: Robustness against statistical null models for nestedness. Unsurprisingly, results for the null model I (much less restrictive) yield extremely high z-scores, as opposed to the comparatively moderate results from null model II (note the y=x line as a visual aid). Despite these differences, both models are highly consistent at quantifying the level of significance for the nestedness.

## C. Modularity

Modularity was originally proposed as a metric for community detection in networks by Newman and Girvan [50] aiming at identifying the mesoscale organization of networks, which reveals many hidden features invisible from a global perspective of the network; informally, modularity (typically labeled $Q$) relies on the detection of densely connected subgraphs: it quantifies the extent to which nodes in a network tend to cluster together, in comparison to the expected distribution of a random counterpart (null model).

One of the interesting aspects of $Q$ is its reliance on the concept of null model, which can be taken as the baseline against which optimization makes sense. This has allowed the original formulation by Newman to be extended to other scenarios, namely directed, weighted or signed (if we pay attention to the features of the links); and bipartite and multiplex networks, beyond the (more common) unipartite networks. The general layout of $Q$ is

$$Q = \frac{1}{2m} \sum_{i}^{N} \sum_{j}^{N} \left( A_{ij} - P_{ij} \right) \delta \left( g_i, g_j \right) , \qquad (4)$$

where $g_i$ represents the module node $i$ belongs to, $A_{ij}$ is the real adjacency matrix of the network, and $P_{ij}$ are the probabilities that an edge linking nodes $i$ and $j$ exists in the null model.

The key point is to define, in this equation, a suitable, adapted null model to confront the real connectivity patterns (as in the mentioned cases). The issue is controversial because even within a type of network different possible null models can be defined. In the bipartite scheme we find two main proposals. We have chosen to work with Barber's definition of modularity for bipartite networks [26] (see also the main text), ruling out the proposal by Guimerà *et al*. [47]. In Guimerà's proposal, modules are forced to be defined strictly in class "purity", that is, a module can only contain nodes of a single class. His method is thus almost equivalent to optimize modularity on the projected unipartite network, which collapses the information in the bipartite original network onto one of its classes (see [47] for the details). On the contrary, Barber's definition naturally incorporates combined (or mixed) modules, formed by nodes from both classes.

The choice of one or another definition is a matter of the problem one intends to solve. Indeed, it may not make a lot of sense to define movie-actor mixed modules, because the semantics of such a module is not very clear. In other problems, however, it may be more convenient to allow for mixed modules. This is often the case in ecology (as for instance in [40]), because it is more interesting to identify modules that have a precise biological meaning as potential co-evolutionary groups [53] or as cores of mutualistic networks [9]. As we are also, in our user-meme systems, more interested in this co-evolutionary perspective, we have taken Barber's approach.

We have applied this metric making use of the software provided by Marquitti *et al*. [49], where the simulated annealing method [27] is used to maximize Barber's modularity. Statistical significance of the results is checked obtaining the z-score of the original network modularity, against the average and standard deviation of 100 random realizations (null model II as for nestedness, see above).

**D. Nestedness and Modularity: further considerations**

**Robustness across window widths**. Beyond assessing the robustness of the results for the nestedness values (regarding the used metrics and null models), we also need to test for robustness against the (admittedly arbitrary) choice of a window-width. This applies both to the soundness of the results in nestedness and modularity.

In Figure S4 we report results for modularity and nestedness (both in their standardised version) for every width $w$ we have tested. Upon inspection, it is clear that results are noisier the narrower the window is –the regularity of the peaks suggests that the measures are sensitive to circadian rhythms (periodic temporal patterns). For values aggregating the activity for one day and beyond, such periodic variations disappear. Remarkably, nestedness (lower panel) shows the same trend for every window width. In contrast, the trajectory of the modularity z-scores is coherent up to $w = 1$ day, but it is blurred out for $w = 3$ days. These results (together with those obtained for the UK dataset) suggest that events have their very own characteristic timescale [55], and observed trends are valid only within a relatively precise range.

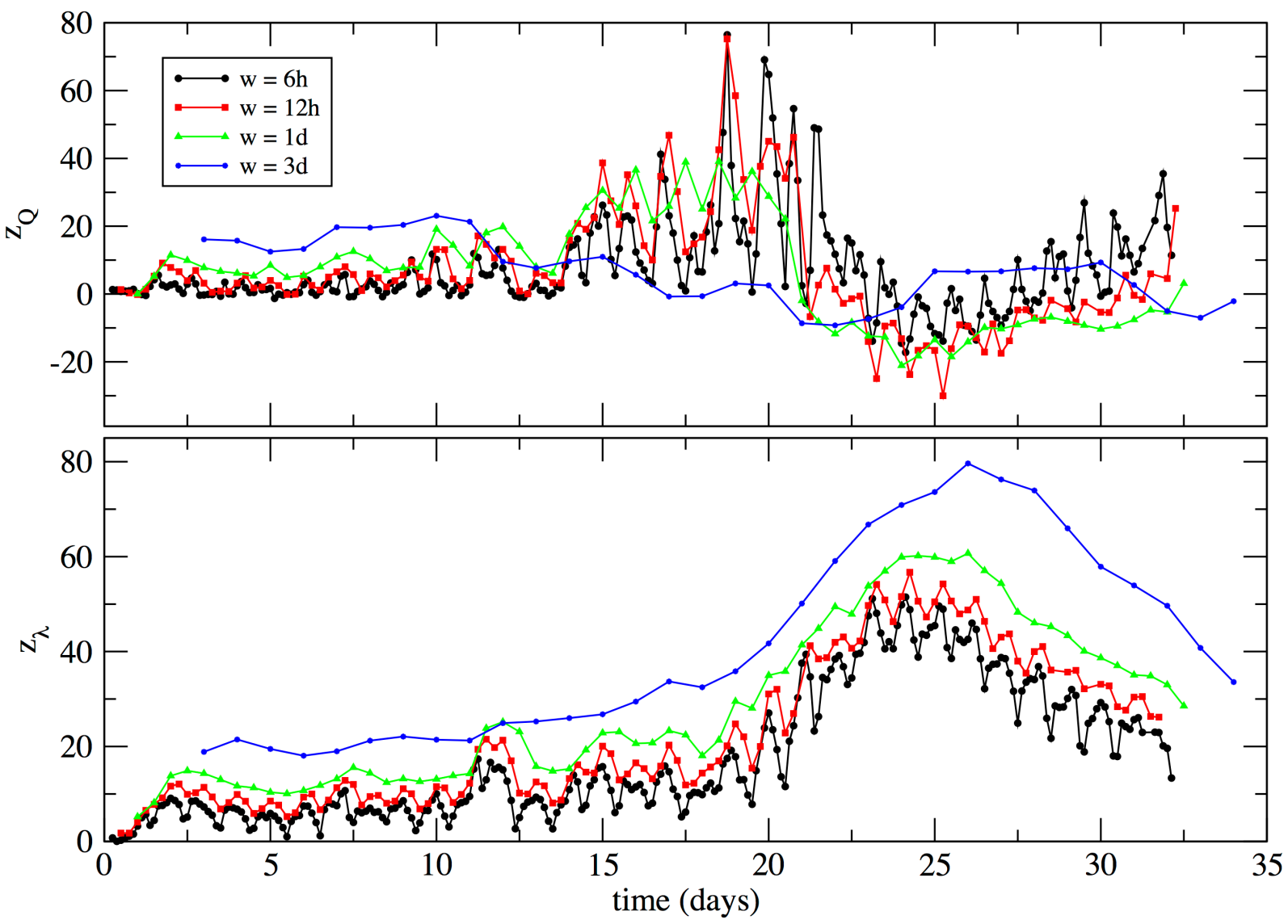

Figure S4: Robustness against window width. Standardised nestedness, $z_{\lambda}$ , and modularity, $z_{Q}$, values are displayed for window sizes 6h, 12h, 24h and 72h.

We observe the same robustness in the UK datasets for a window width of $w$ = 2 hours (as opposed to $w$ = 1 hour reported in the main text). Given the fast time scale of the event (it fully develops in less than two days), wider window schemes blur the results.

**Ruling out epiphenomenal conclusions**. Both in the main text and throughout this document we have provided solid evidence that, in an information ecosystem such as Twitter, topics arise in a nested scheme out of an initially modular structure. One may argue, however, that this striking outcome may be artificial in different senses. First, it is possible that the modular-to-nested transition occurs out of a "topological artifact", namely, that the network starts as a broken set of small components (thus being trivially modular) and undergoes a percolation process such that nestedness is possible from then on. In Figure S5 the size of the giant component of the system is plotted as it evolves in time. Such component is always above $0.78N$, and as such a percolation transition is never observed.

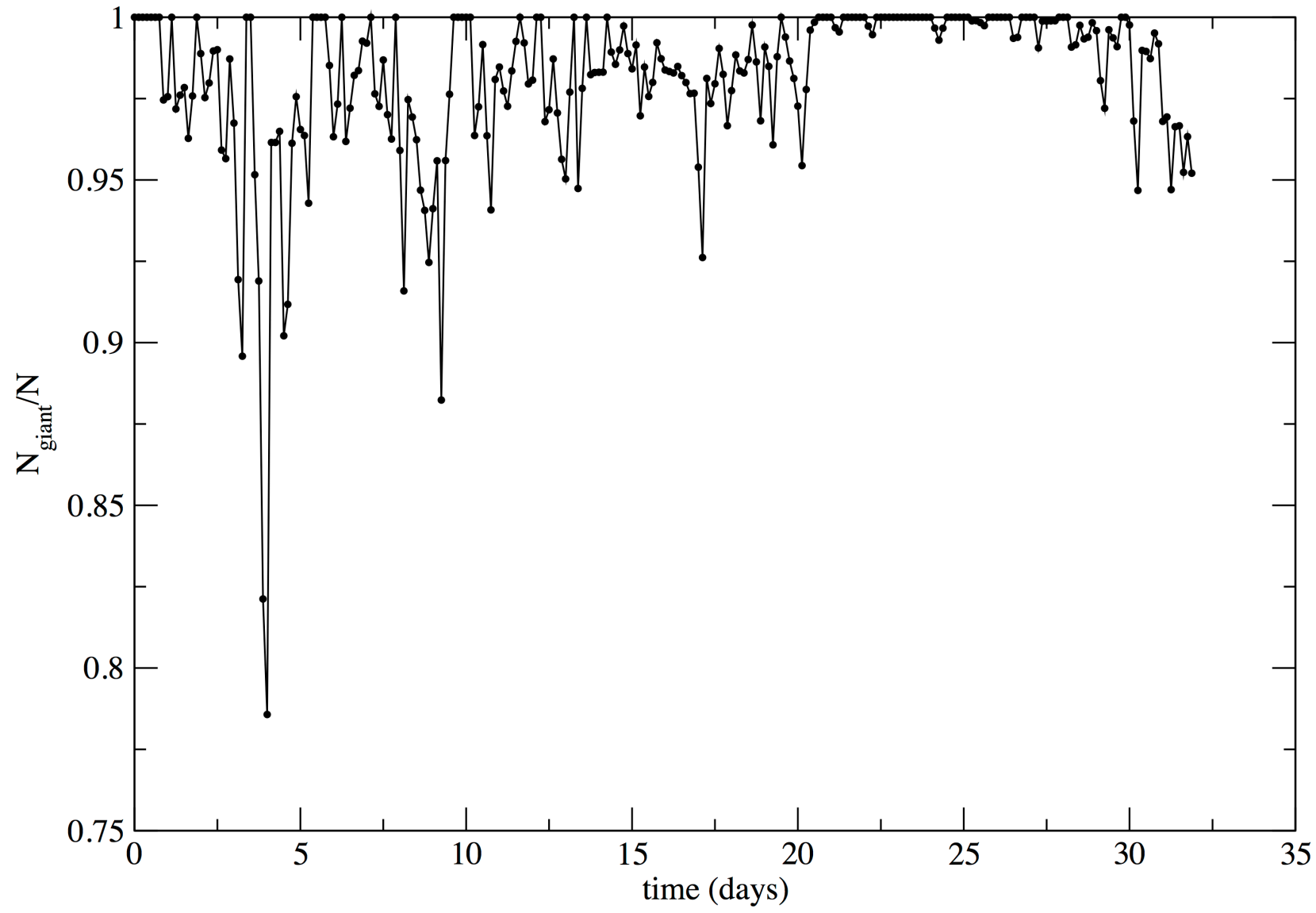

Figure S5: Evolution of the size of the giant connected component (as a proportion of the total size of the system). Notably, the *y*-axis is labeled from 0.75 and above, which implies that, for the whole time range (over a month), the system does not undergo any abrupt percolation process. The figure corresponds to a window width $w = 6$ hours (the noisiest and sparsest one).

A second consideration implies our disregarding of weighted values. Indeed, we have focused on binary, presence/absence matrices –in an effort to follow the ecosystems literature. Additionally, NODF does not have, to our knowledge, a weighted equivalent, so comparison is properly done only with a binary representation of the system.

Admittedly, this represents a loss of information, which could potentially affect the results. We are aware that the spectral radius approach to nestedness does allow for weights to be present in the interaction matrix. For the sake of completeness, we have measured nestedness also considering weights, which stand for the frequency with which an individual used a certain hashtag, given a certain time window. The result can be seen in Figure S6, where the growing pattern follows precisely the trends reported in the main text (Figure 2) and here (Figure S4).

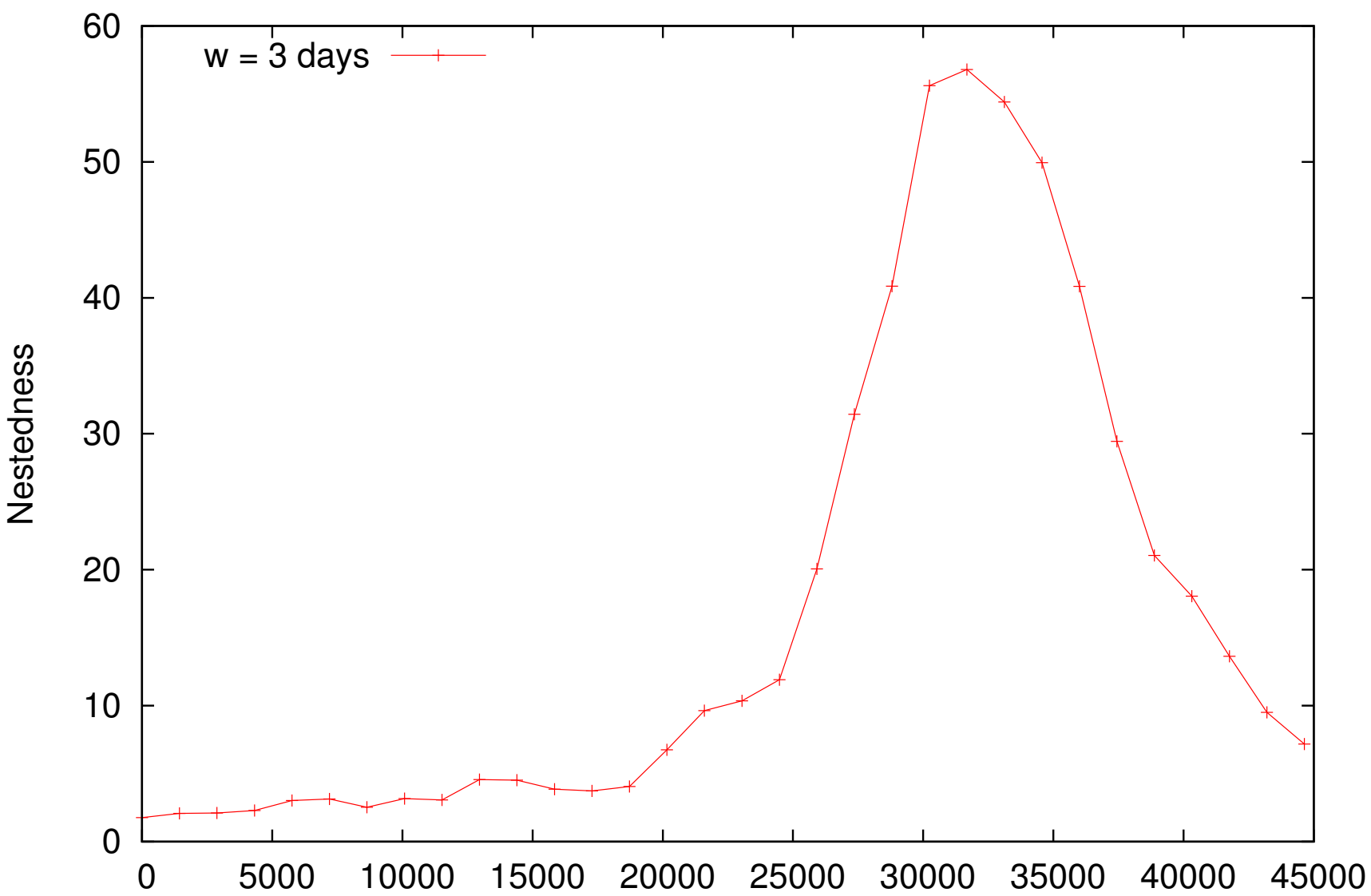

Figure S6: Nestedness evolution as measured from weighted matrices, i.e. matrices which encode the absolute usage counts of hashtags by the corresponding Twitter users. The figure corresponds to a window width $w$ = 3, and delivers the same growing patterns as its own binary counterpart. (time in the x-axis is expressed here in minutes since the origin of our data; here, $t$ = 30000 is roughly equivalent to May15th).

## E. Mutualistic Dynamical Model

According to the dynamical framework from Bastolla *et al*. [36], the evolution of a mutualistic ecosystem can be modeled through a set of $N$ differential equations in which each equation represents the variation of the frequency of a plant or a pollinator. Competitive interactions are estimated through linear function responses, $-\beta_{ij}^{(P)}N_j^{(P)}$ for plants and $-\beta_{kl}^{(A)}\, N_l^{(A)}$ for animals, where competition matrices $\beta_{ij}^{(P,A)}$ are symmetric and non-negative. In the same way, mutualistic interactions between plants and pollinators are modeled through non-linear functional responses of Holling Type, $f(N) = \gamma N/(1+h\gamma N)$, where mutualistic relationships are described through two symmetric and non-negative matrices $\gamma_{ij}^{(P,A)}$ and the Holling term $h$ imposes a limit to the mutualistic growth rate, avoiding divergences in the case of large populations. The equations for the system's dynamics are described in *Materials and Methods* section of the main text.

### E.1. Synthetic topologies

Regarding topologies, we built *ad hoc* two ensembles of synthetic networks for different network sizes. For each pair of ensembles (equal size, equal link density), the networks of the first ensemble present an almost perfectly nested architecture, while the networks of the second one exhibit an almost perfectly modular structure. All the networks of a given size $N$ were built with the same number of users and hashtags $n = m$. Nested networks were constructed starting from a perfect nested structure, involving a connectivity distribution $k_u = u$, $u = 1, 2, \ldots$; $k_h = h$, $h = 1,2,\ldots$, and subsequently randomizing each link with probability $p = 0.02$. This method provides networks with an almost perfect nested topology and mean connectivity $<k> = N/4$.

According to the procedure by Newman [50], modular networks were constructed starting from a perfect modular structure consisting of 5 disconnected blocks (cliques) of equal size $N_i = N/5$, and subsequently randomly connecting pairs to reach the connectivity $<k>= N/4$. The number of blocks (5) and the rewiring probability $p = 0.02$ were arbitrarily chosen. Nevertheless, the results are robust against variations of these values, as shown in figure S7.

### E.2. Realizations

In each realization, we used a different network of the corresponding ensemble, and assigned different random initial frequencies to each user and hashtag in the interval $s_{u,h}(t = 0) \in (0, 1)$, and different growing rates in the interval $\alpha_{u,h} \in (0.85, 1.1)$. We ran the dynamics defined by equations (4) and (6) of the main text and, once the stationary state was reached, we computed the survival rate by adding the number of users and hashtags with frequency $s_{u,h} > 0$ and then divided by their initial number $N$. Accordingly, the survival area stands for the region of the parameters space with a survival rate greater than a given value. We performed 1000 different realizations per each point of the space of parameters $\beta \times \gamma$ and for each size and topology. According to the standard biology procedures (see, e.g., [36, 53]), the inter-specific term was fixed to $\rho = 0.2$ and the Holling term was set to $\lambda = 0.1$. The values for the competition and mutualistic terms covered the range $\beta,\gamma \in [0.1]$ with intervals of $\delta_\beta$ , $\delta_\gamma = 0.05$ (from weak to strong mutualism regimes).

Results of these extensive simulations are shown in Figure 3 of the main text, where left panels represent the survival rate (i.e., the diversity in the stationary state) as a function of the competitive and mutualistic terms $\beta$ and $\gamma$, for a system size of $N = 1000$. Right panels of Figure 3 of the main text represent the normalized area of the space of parameters $\beta \times \gamma$ that exhibits a survival rate equal o higher than the corresponding value of the $x$-axis, for different network sizes in different panels: $N = 50$, 100, 300, 1000. As discussed in the main text, a large area of the space of parameters exhibits high persistence for the nested architecture where persistence is low for the modular one, *but never the opposite*. Otherwise, for the modular architecture, the area of the space $\beta \times \gamma$ with a given persistence decreases sharply with the network size, while for the nested architecture this dependence is smaller. This effect saturates for large values of the network size, that is, once the size $N \sim 500$ is reached, the size of the network does not have a noticeable effect on persistence anymore. Figures S8 and S9 complement, respectively, panel left and right of Figure 3 of the main text. Figure S8 represents the persistence (i.e. diversity of memes and hashtags in the stationary state) for each value of $\beta$ and $\gamma$, for nested and modular networks of 100, 500, and 1000 nodes. Additionally, Figure S9 represents the normalized area of the surface ($\beta \times \gamma$) that exhibits a persistence equal o higher than a given value as a function of that value, for nested and modular networks. In the above results (Figure 3 of the main text and Figures S6-S9), the interval $\alpha_{u,h} \in (0.85, 1.1)$ has been taken according to the biological literature (see, e.g. [36]). Nevertheless, the main result (nested architectures out-survive modular ones) holds for wider interval of $\alpha_{u,h,}$ as shown in figure S11 for $\alpha_{u,h} \in (0.5, 1.5)$.

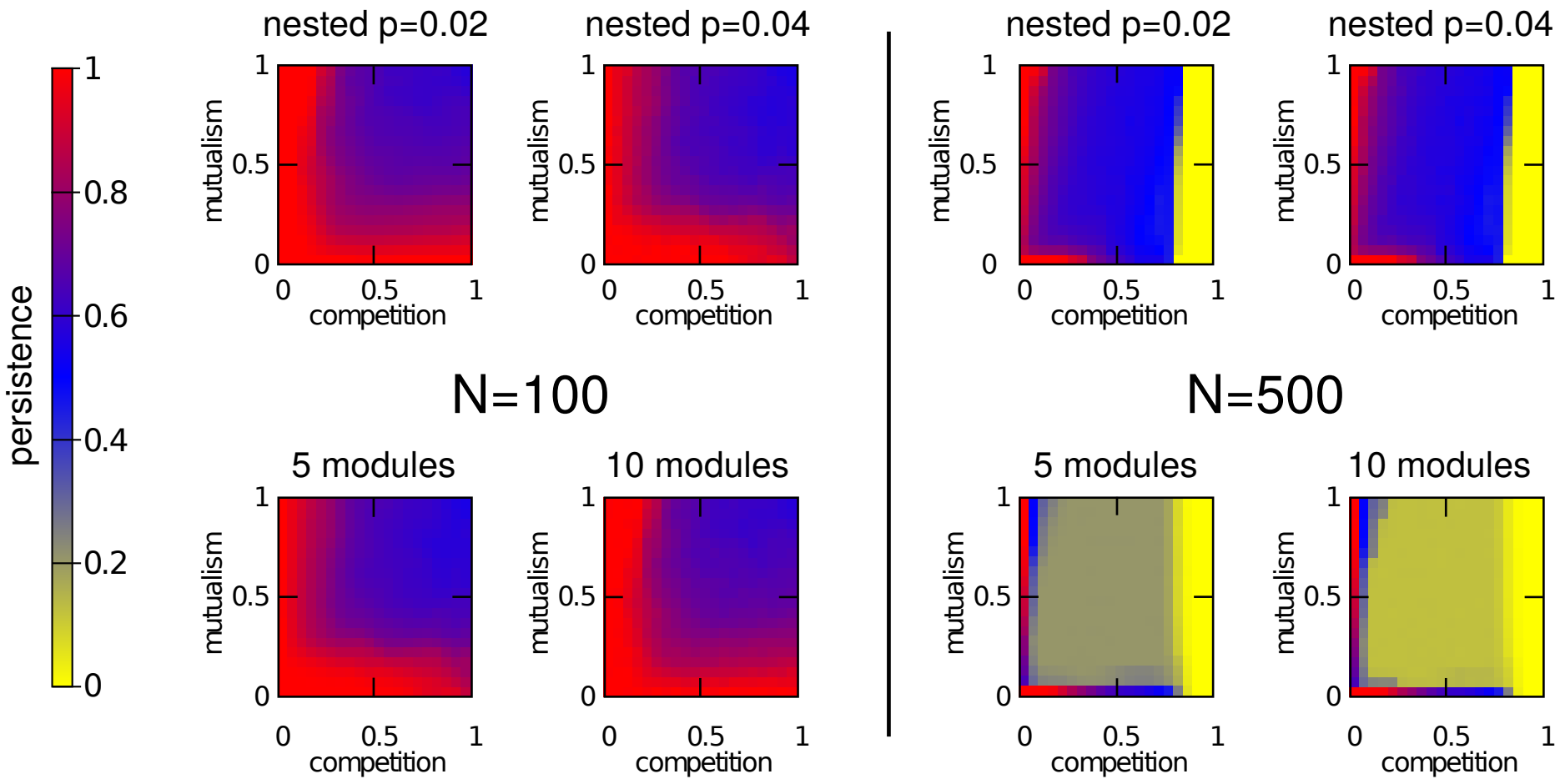

Figure S7: Survival rate as a function of $\beta$ and $\gamma$, for different values of the rewiring probability $p$ in the nested networks (up panels) and different number of modules in the modular networks (down panels). Left panels correspond to a network size of $N$ = 100, while right panels correspond to $N$ = 500. Each point is averaged over 1000 different initial conditions.

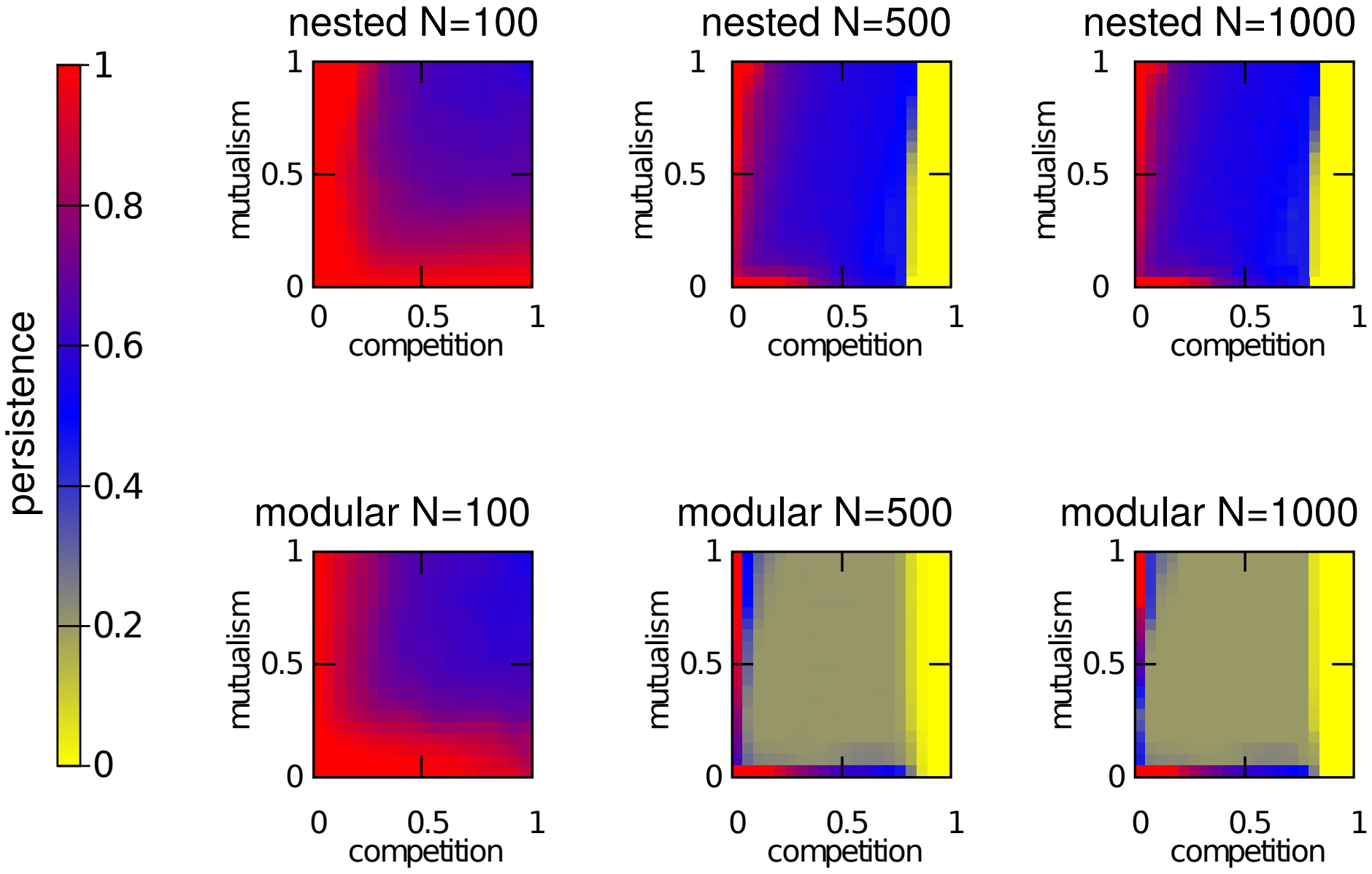

Figure S8: Survival rate as a function of the competition $\beta$ and mutualism $\gamma$ terms. Upper (resp., lower) panels correspond to a nested (resp., modular) architecture. Each column corresponds to a different value of the network size: $N$ = 100 (left), $N$ = 500 (center), $N$ = 1000 (right). Each point is averaged over 1000 different initial conditions.

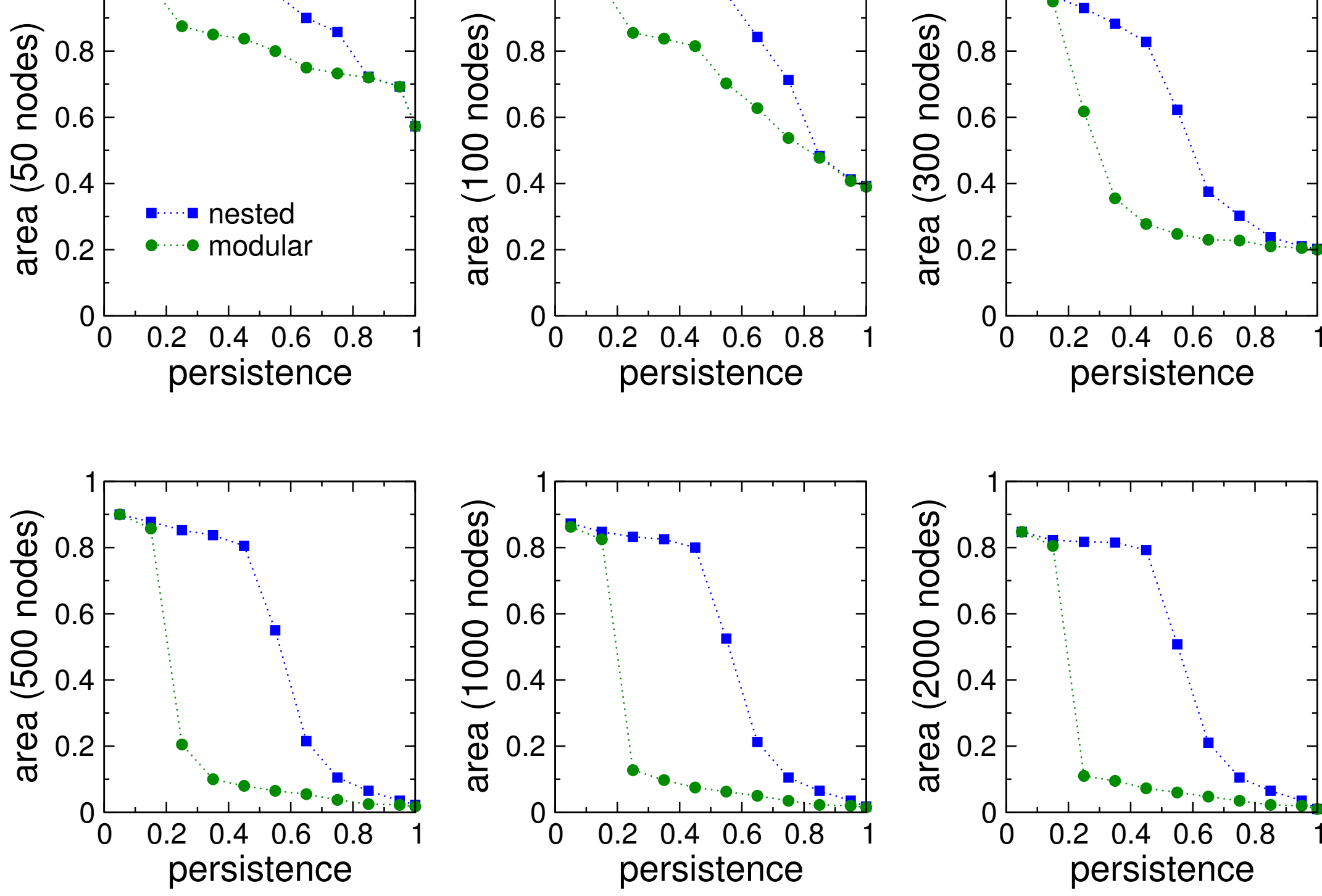

Figure S9: Normalized area of the space $\beta \times \gamma$ with a survival rate equal o higher than the corresponding value of the $x$-axis. Different panels correspond to different values of the network size: $N$ = 50, 100, 300, 500, 1000, 2000. Note that, for the sake of comparison, 4 of the panels are merely reproducing those reported in Figure 3 of the main text.

The value $\rho$ = 0.2 in the simulations from Figure 3 is taken from the ecological literature, where the intra-species competitive term (1 - $\rho$) is usually considered to be greater than the inter-species term ($\rho$), as members of the same species are competing for the same resources. For the sake of completeness, we have studied as well the case N = 1000, $\rho$ = 0.6, which in our study corresponds to an inter-users stress greater than the intra-users stress. Figure S10 shows that, when the inter-users stress exceeds the intra-users stress, the main feature observed in the dynamics remains intact, namely, modular networks exhibit poor survival, while nested networks show equal or higher levels than the modular structure in any given region.

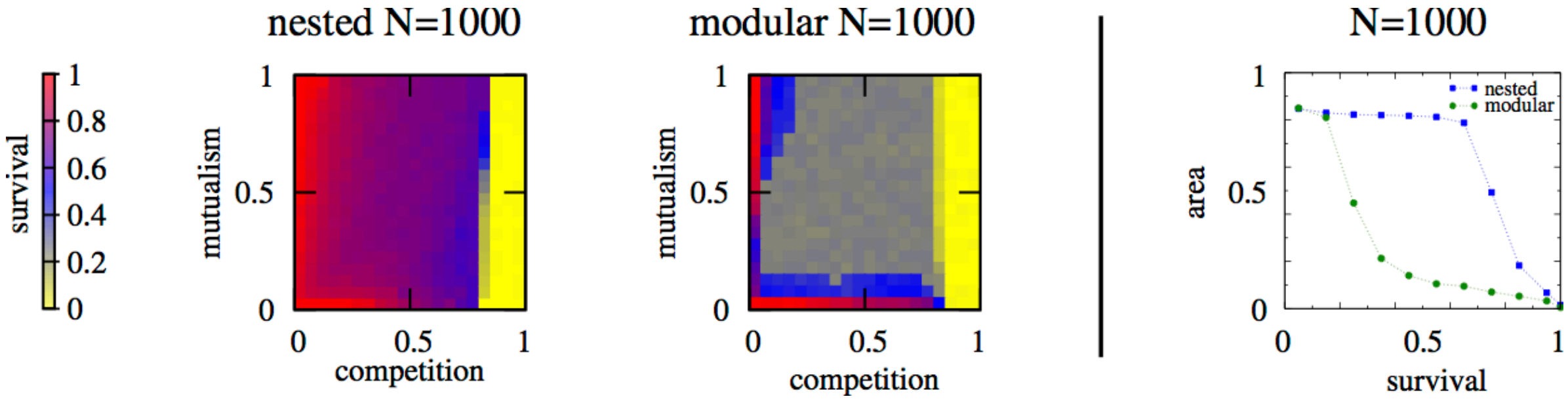

Figure S10: Survival rate (left) and normalized area survival (right) for a network with $N$=1000 and $\rho$ = 0.6 (corresponding to Figure 3 of the MT, and Figures S8 and S9 of the Supplementary Materials).

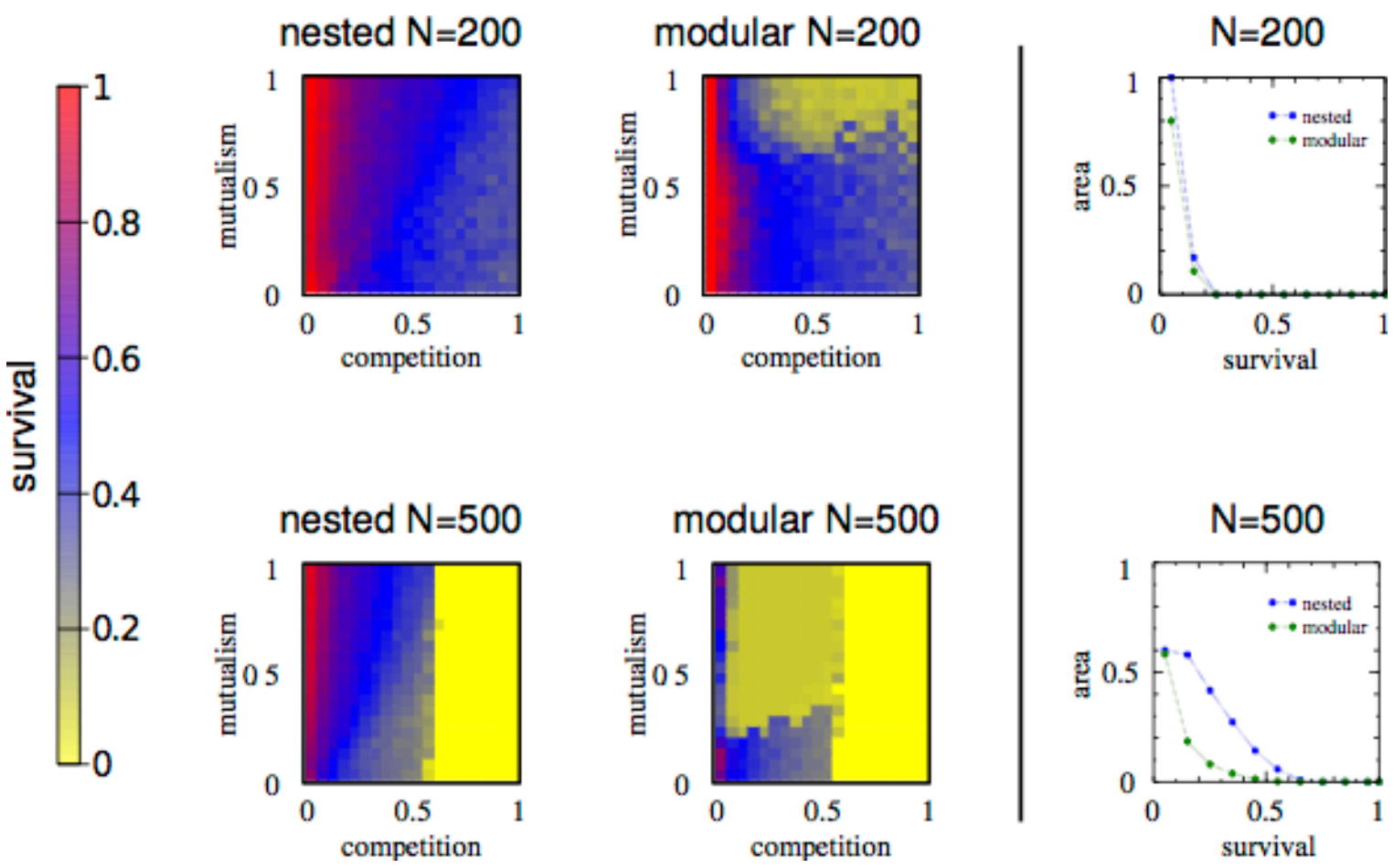

Figure S11: Left and center panels show the survival rate (color code) as a function of competition (x-axis) and mutualism (y-axis). Left (resp., center) panels correspond to a nested (resp., modular) architecture. Different raws correspond to different values of the network size: $N$ = 200, 500. Right panels show the normalized area of the space $\beta \times \gamma$ with a survival rate equal o higher than the corresponding value of the $x$-axis. $\alpha_{u,h} \in (0.5, 1.5)$.

## F. Core-periphery structure

Meso-scale structures in networks have received considerable attention in recent years, as the detection of these intermediate-scale patterns can reveal important characteristics that are hidden at both local and global scales. Among the wide diversity of methods aiming at the detection of such structures, community detection methods have become very popular and successful. In this section we focus our attention on a different type of meso-scale structure, known as the *core-periphery* structure, that helps one to visualize which nodes of the graph belong to a densely connected component or *core*, and which of them are part of the network's sparsely connected *periphery*. Nodes belonging to the core should be relatively well connected to other nodes in the network, either central or peripheral; whereas nodes in the periphery should be those elements poorly connected with the core, and disconnected from the periphery. According to this intuitive notion, many methods have been proposed. We follow here a method developed by Della Rossa *et al*. [38], based on the profile derived by a standard random walk model. It and can be obtained in a very general framework and is applied here for undirected unweighted networks.

Let $w_{ij} = w_{ji}$ be the link of weight 1 between nodes $i \leftrightarrow j$ in our network of size $N$. At each time step, the probability that the random walker at node $i$ jumps to node $j$ is given by $m_{ij}$:

$$m_{ij} = \frac{w_{ij}}{\sum_h w_{ih}} = \frac{1}{k_i} \qquad (5)$$

where $k_i$ is the degree of $i$. The asymptotic probability of visiting node $i$ has the closed form

$$\pi_i = \frac{k_i}{\sum_j k_j} \qquad (6)$$

The method starts by randomly selecting a node $i$ among those with the weakest connectivities, and assigning $\alpha_i = 0$. Pk, the set of nodes that are already assigned at step $k$, is then filled with $i$, $P_1 = \{i\}$. For the following steps, $k = 2, 3, ..., n$, the node $j$ attaining the minimum in

$$\begin{aligned}\alpha_j &= \min_{h \in N \setminus P_{k-1}} \frac{\sum_{i,j \in P_{k-1} \cup \{h\}} \pi_i m_{ij}}{\sum_{i \in P_{k-1} \cup \{h\}} \pi_i} \\ &= \min_{h \in N \setminus P_{k-1}} \frac{\sum_{i,j \in P_{k-1}} \pi_i m_{ij} + \sum_{i \in P_{k-1}} (\pi_i m_{ih} + \pi_h m_{hi})}{\sum_{i \in P_{k-1}} \pi_i + \pi_h}\end{aligned} \qquad (7)$$

is selected. If it is not unique, a randomly chosen node among them, $l$, is selected, and $P_k = P_{k-1} \cup l$. Although the algorithm presents some randomness, it has been verified that the effect in the analysis of real-world networks is negligible. The core-periphery profile is then the set $\{\alpha_k\}$, with $0 \leq \alpha_k \leq 1$,where $\alpha_k = 0$ for nodes belonging to the periphery and $\alpha_k > 0$ for nodes in the core.

As the goal of this section is to identify the possible formation of a core during the days preceding the 15M, a distance metric should be defined. As a first approach we consider the distance between two core-periphery structures as the product between the two $\{\alpha_k\}$ sequences,

$$\vec{\alpha}_1 \cdot \vec{\alpha}_2 = \sum_{i=0}^{n \times m-1} \alpha_{1i} \alpha_{2i} \qquad (8)$$

Notice that the $\alpha$ "vectors" do not necessarily share the same coordinates, that is, it may happen that a given node (now by nodes we refer to users or hashtags indifferently) present in $\vec{\alpha}_1$ does not appear in $\vec{\alpha}_2$ because it was not part of the network. Whenever this is the case, we consider the contribution to the dot product to be zero (i.e., as if it were at the periphery). On the other hand, we normalize the above expression in order to get a bounded value: $0 < \vec{\alpha}_1 \cdot \vec{\alpha}_2 < 1$ ,

$$\vec{\alpha}_1 \cdot \vec{\alpha}_2 = \frac{1}{\|\alpha_1\| \|\alpha_2\|} \sum_{i=0}^{n \times m-1} \alpha_{1i} \alpha_{2i} \qquad (9)$$

which is the expression used in the main text and labeled as $D_{RC}$.

In the main text we have discussed the conformation of a relatively stable core of hashtags around the 15M day, in contrast to a high turnover of users coming to and leaving the core at different snapshots. Here, we scrutinize further such a finding by ruling out the possibility that it could be due, for example, to the fact that the set of users in the core could be similar over the distinct time-windows and change abruptly at the reference point under consideration. To this aim, we additionally measured the distance of a given core $C_t$ from the previous core $C_{t-1}$ –the core present in the previous time-stamp. Results in Figure S12 reject this conjecture: the turnover of users is still high –the distance is small– when the core is compared with that in the preceding graph, suggesting that users are actually entering and exiting the key positions in the network. In contrast, hashtags keep relatively constant at high distances, indicating that the core near the 15M is formed smoothly –the exception to this being a sharp decrease around the 15M if observed at a 3-day window resolution. The reason for this behavior is the takeover of new hashtags (with respect to the ones that originated the protest), which pushes the original ones away from the core of the structure. The best example of this is the hashtag *#democraciarealya* ("real democracy now"), which is placed at the core of the bipartite network for a long period of time, but its leading role is substituted by the more generic (and "cheap" from a microblogging perspective) *#15m* from the immediately previous days of May 15 and onwards.

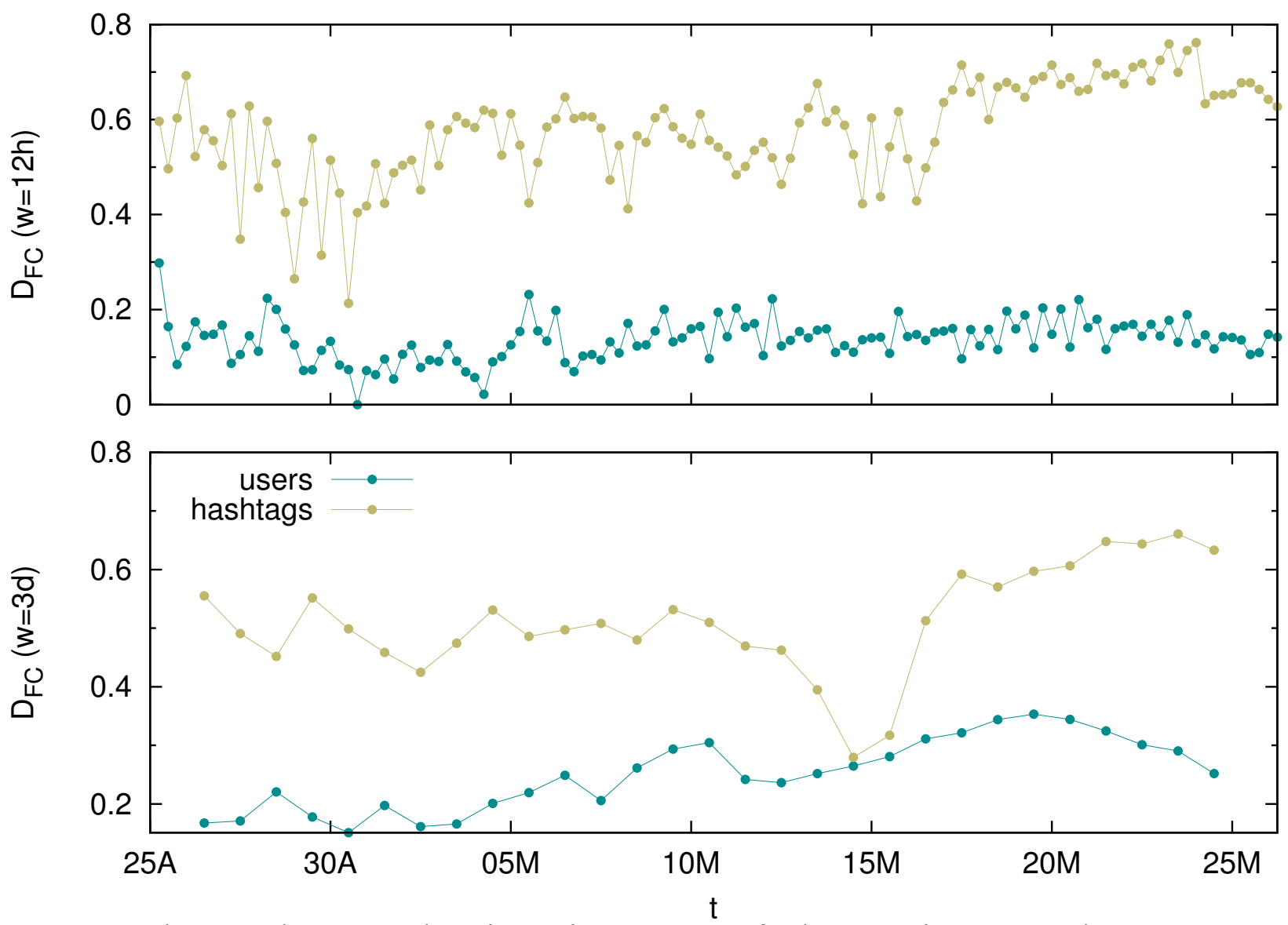

Figure S12: The figure emulates the results in Figure 3 of the main text; however, distances are computed between each time stamped core and the former configuration, $C_t$ vs. $C_{t-1}$. As in the main text, the figure illustrates that hashtags build a more stable core in comparison to users.

## G. Anti-correlation between nestedness and modularity

Further details on the reported $Q$-nestedness correlated/anti-correlated patterns can be seen in Table S2 and Figure S13.

| Dataset | | $r$ | | $p$-value |
|---|---|---|---|---|
| 15M | $w$ = 6h | pre May 15 | 0.8182 | $10^{-5}$ |
| | | post May 15 | -0.7569 | $10^{-5}$ |
| | $w$ = 12 h | pre May 15 | 0.8179 | $10^{-5}$ |
| | | post May 15 | -0.8023 | $10^{-5}$ |
| | $w$ = 24 h | pre May 15 | 0.7997 | $10^{-4}$ |
| | | post May 15 | -0.7819 | $10^{-5}$ |
| | $w$ = 72 h | pre May 15 | 0.5438 | not significant |
| | | post May 15 | -0.2732 | not significant |
| UK | $w$ = 1h | –0.7126 | | $10^{-5}$ |

Table S2: Pearson correlation coefficient values for nestedness and modularity, along with their $p$-values, for both datasets and available window widths. In the case of 15M, we report Pearson correlations before and after the climax of the event (though the exact moment at which modularity abruptly collapses is slightly different in each case, see panels in Figure S13). Note that correlations fail to be significant (that is, $p > 0.05$) for $w$ = 3 days, as the results for modularity are blurred compared to the observed pattern in $6h \leq w \leq 24h$.

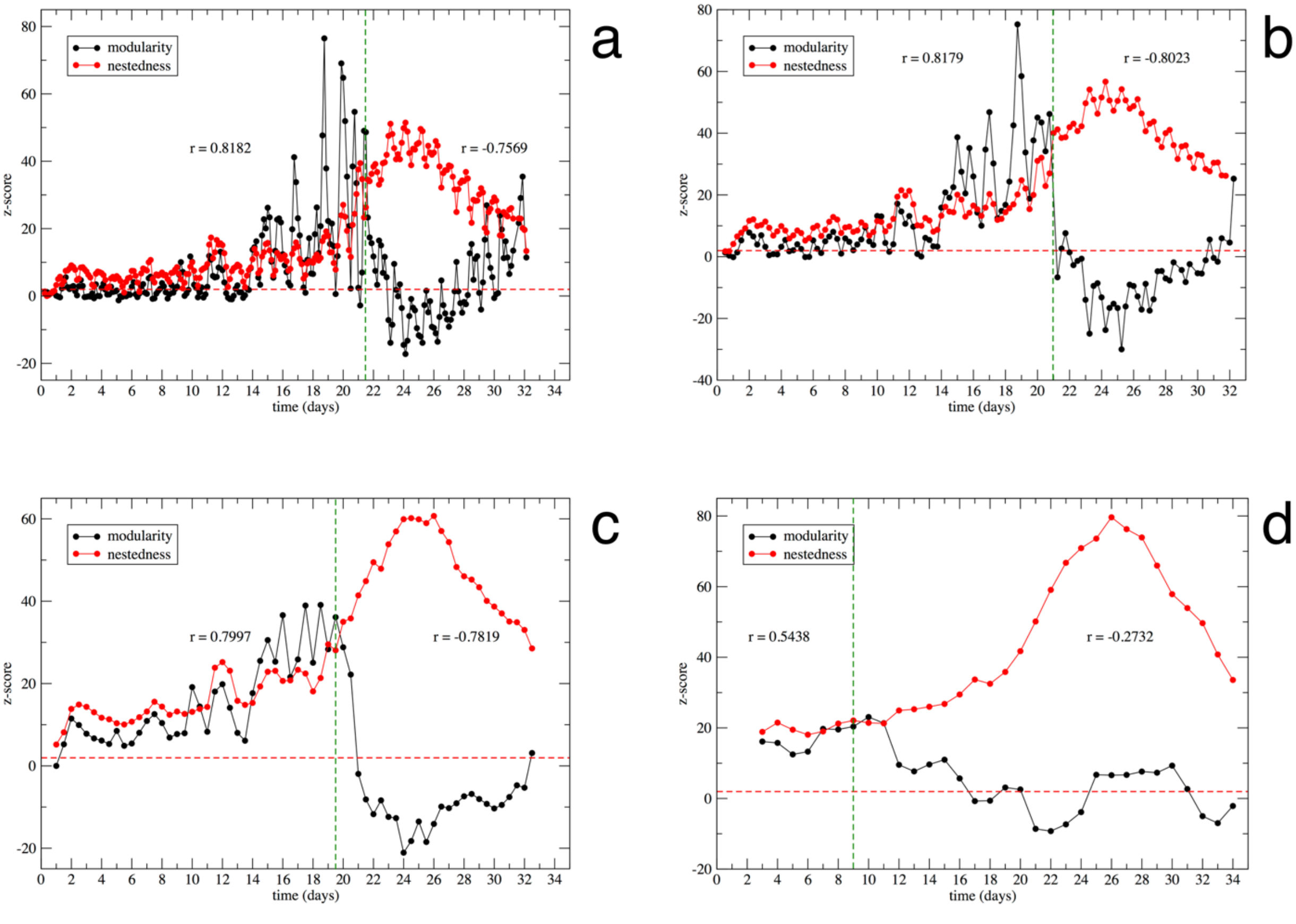

Figure S13. Evolution in time of nestedness and modularity z-scores for all the window widths (6h, 12h, 24h and 72h) in the 15M dataset (c panel is the same on as reported in the main text, and repeated here for the sake of comparison). Around the onset of the 15M protests modularity collapses and displays an anti-correlated behavior with respect to nestedness. An exception to this is $w$ = 3 days (panel d), in which the aggregated data in a single snapshot blurs the results of the community detection algorithm. A vertical green line has been placed on the time when nestedness and modularity bifurcate, and Pearson coefficients are reported for both sides of that line. A red horizontal line has been placed at z = 1.96, as a visual aid for statistical significance.

**H. "Topic" null model**

We have discussed in the beginning of this supplementary information distinct possibilities regarding the construction of presence-absence matrices that describe the set of interactions in our systems. We have also mentioned that different cutoffs to hashtags and/or users can be applied, and discussed the more reasonable way to proceed to study nestedness and modularity, which consists of either considering the most active users and their related set of hashtags, or the set of most active users plus most tweeted hashtags at a time. Also, on the side of statistical soundness, we have delved into different null model possibilities.

Now however, our concern focuses on the singularity of the results themselves. In particular, we want to test whether the modularity-nestedness crossover we have observed for particular topics is universal to *any* activity on Twitter (and in this sense uninteresting), or rather it is a specific mechanism underlying the formation of consensus around related information. Thus we explore here three additional possibilities for the $w$ = 12h time-window on the UK dataset. In option (a) we select randomly and independently 512 users and 512 hashtags, and build the corresponding presence-absence matrix. Although the way in which nodes are selected can produce empty matrices corresponding to graphs with no links, this never happened in our dataset (all matrices have more than 20 non-empty cells). In model (b), 512 users are randomly selected and they determine the set of hashtags to consider. Model (c) is analogous but selecting randomly the 512 hashtags to be included, along with the set of users that tweeted them.

These three sets, (a), (b) and (c), can be considered as an additional category of null models that allow us to discern if the nested patterns previously observed are significant: for example, if set (a) showed high levels of $z_\lambda$ we would not be able to conclude that the coordination phase observed in the 15M is relevant, as we would be finding nested patterns even for structures randomly filtered. A comparison between the three methods is displayed in Figure S14. Results include data from Figure 5 (bottom panel) in the main text. We observe that, when we consider independent users and hashtags at random –set (a)– nested patters do not show up and the bipartite network do not present any kind of organized structure. The exception is the region between the 3rd of February afternoon and the 4th of February, when the XLVII Super Bowl took place, probably due to the high relevance of this tournament (if it became global trending topic, even a randomly built network would show, to some extent, a nested structure). When users (hashtags) are randomly selected, but the set of hashtags (users) is closely related to them, the nestedness increase –sets (b) and (c)–, but this is a systematic shift rather than a differential change.

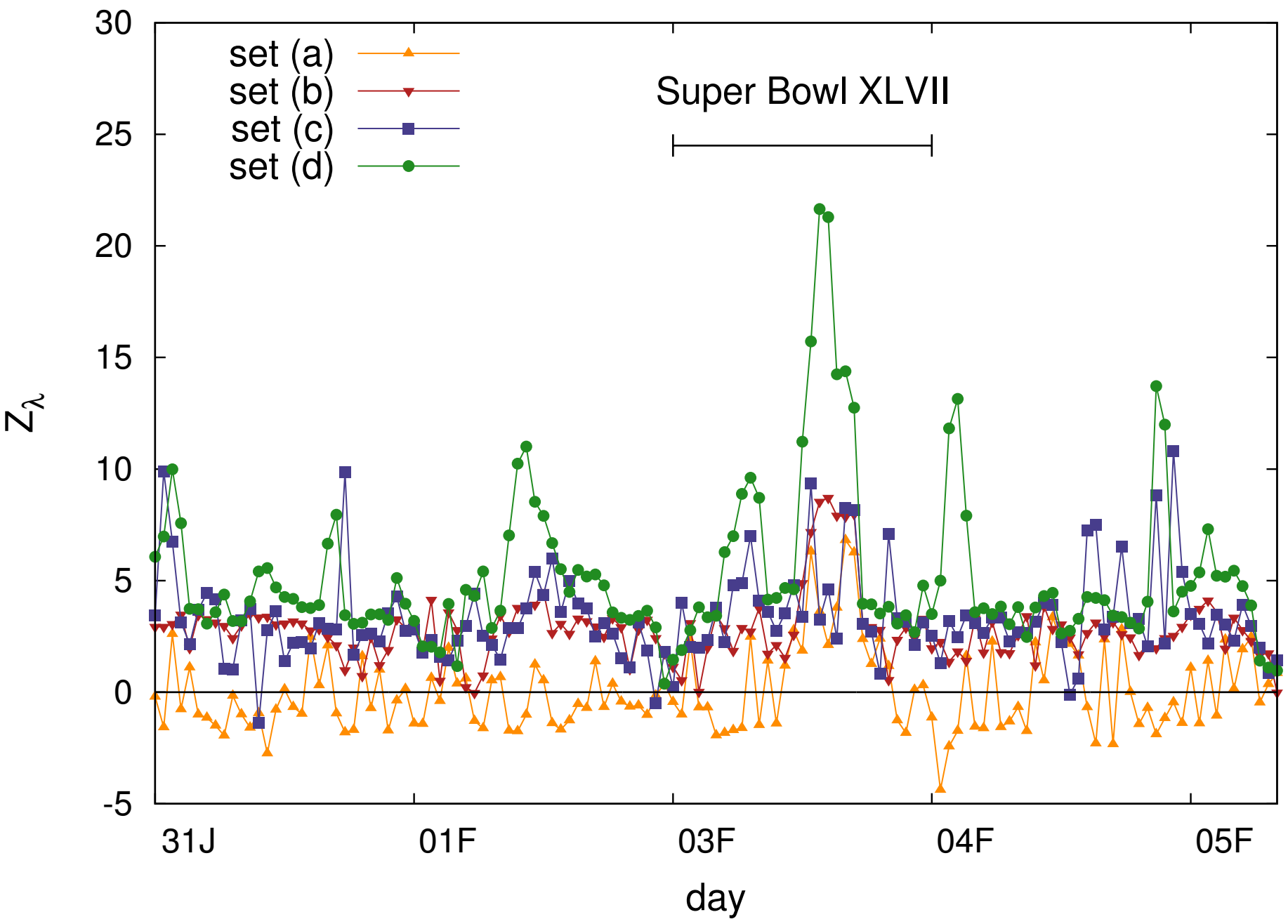

Figure S14: Some results for unfiltered Twitter traffic (2013). Set (a) corresponds to our UK dataset with 512 hashtags and 512 users randomly and independently selected. Such a random selection implies that the presence-absence matrix might be empty, although it never happened in this case. In set (b), 512 users have been randomly chosen determining the set of hashtags. Inversely, set (c) have been obtained by randomly filtering 512 hashtags their related users. Finally, set (4) comprises the 512 most active users and the 512 most used hashtags for comparison.

## Appendix. Some selected hashtags

In Tables S3 and S4 we display some of the hashtags used in our dataset, along with the number of counts registered.

| Hashtag | Counts | Hashtag | Counts |
|---|---|---|---|
| #acampadasol | 189251 | #acampadazgz | 6033 |
| #spanishrevolution | 158487 | #acamapadasol | 5760 |
| #nolesvotes | 66329 | #22m | 5205 |
| #15m | 65962 | #reflexion | 4693 |
| #nonosvamos | 55245 | #sinbanderas | 4481 |
| #democraciarealya | 47463 | #consensodeminimos | 4348 |
| #notenemosmiedo | 32586 | #italianrevolution | 3981 |
| #yeswecamp | 31811 | #estonosepara | 3860 |
| #acampadabcn | 20069 | #acampadaalicante | 3593 |
| #15mani | 17986 | #tomalacalle | 3517 |
| #acampadasevill1a | 14356 | #fb | 3372 |
| #globalcamp | 13186 | #europeanrevolution | 3035 |
| #acampadavalencia | 13129 | #acampadapamplona | 2839 |
| #estoesreflexion | 11080 | #worldrevolution | 2777 |
| #acampadagranada | 9717 | #democraciarealya | 2766 |
| #acampadamalaga | 6808 | #acampadapalma | 2709 |

Table S3. Top 32 most-used hashtags in the 15M dataset.

| Hashtag | Counts | Hashtag | Counts |
|---|---|---|---|
| #london | 127082 | #essex | 38043 |
| #ff | 110972 | #cbb | 37741 |
| #uk | 80238 | #lol | 37646 |
| #love | 70618 | #tired | 37316 |
| #jobs | 70268 | #legend | 37225 |
| #weather | 66366 | #summer | 36739 |
| #excited | 64795 | #cute | 33648 |
| #mufc | 60176 | #bgt | 33586 |
| #endomondo | 55537 | #coys | 33380 |
| #xfactor | 51870 | #wimbledon | 32510 |
| #lfc | 47237 | #nufc | 32328 |
| #stalbans | 45117 | #help | 32118 |
| #mtvhottest | 43572 | #amazing | 30124 |
| #happy | 40086 | #bored | 30066 |
| #loveit | 40049 | #bbuk | 29960 |
| #nowplaying | 39798 | #ukweather | 29940 |

Table S4. Top 32 most-used hashtags in the UK dataset.